\documentclass[reprint,superscriptaddress,tightenlines,aps,pre]{revtex4-2}
\usepackage{microtype} % improves general appearance of the text
\usepackage{graphicx,epsfig}% Include figure files
\usepackage{dcolumn}
\usepackage{bm}
\usepackage[utf8]{inputenc}
\usepackage{amsmath, amsthm, amssymb, amsfonts}
\usepackage{accents}
\usepackage[usenames,dvipsnames]{xcolor}
\usepackage{tikz}
\usepackage{array}
\usepackage{dsfont}
\usepackage{mathtools}
\usepackage{soul}
\usepackage[multidot]{grffile}
\usepackage[colorlinks=true,linkcolor=Blue,citecolor=Blue,urlcolor=Blue]{hyperref}
\usepackage[caption=false]{subfig}
\usepackage[capitalise]{cleveref}
\usepackage{times}

\newcommand{\epsAB}{\epsilon_{AB}}
\newcommand{\epsBA}{\epsilon_{BA}}

\newcommand{\revision}[1]{\textcolor{black}{#1}}

\setstcolor{blue}   

\begin{document}
\title{Simplicially driven simple contagion}

\author{Maxime Lucas}
\thanks{I.I. and M.L. contributed equally to this work.}
\affiliation{Mathematics and Complex Systems Research Area, ISI Foundation, Via Chisola 5, 10126 Turin, Italy}
\affiliation{CENTAI, Corso Inghilterra 3, 10138 Turin, Italy}
\author{Iacopo Iacopini}
\thanks{I.I. and M.L. contributed equally to this work.}
\affiliation{Department of Network and Data Science, Central European University, 1100 Vienna, Austria}
\author{Thomas Robiglio}
\affiliation{Department of Physics, University of Turin, Via Pietro Giuria 1, 10125 Turin, Italy}
\author{Alain Barrat}
\thanks{A.B and G.P. jointly supervised this work.}
\affiliation{Aix Marseille Univ, Universit\'e de Toulon, CNRS, CPT, Marseille, 13009, France}
\author{Giovanni Petri}
\thanks{A.B and G.P. jointly supervised this work.}
\affiliation{Mathematics and Complex Systems Research Area, ISI Foundation, Via Chisola 5, 10126 Turin, Italy}
\affiliation{CENTAI, Corso Inghilterra 3, 10138 Turin, Italy}

%\date{\today} 
%%%%%%%%%%%%%%%%%%%%%%%%%%%%%%%%%%%%%%%%%%%%%%%%%%%%%%%%%%%%%%%
\begin{abstract} 
Single contagion processes are known to display a continuous transition from an epidemic-free state to an epidemic one, for contagion rates above a critical threshold. 
This transition can become discontinuous when two simple contagion processes are coupled in a bi-directional symmetric way. 
However, in many cases, the coupling is not symmetric and the nature of the processes can differ.
For example, \revision{risky} social behaviors---such as \revision{not wearing masks or engaging in large gatherings}---can affect the spread of a disease, and their adoption dynamics via social reinforcement mechanisms are \revised{usually} better described by complex contagion models rather than by simple contagions, more appropriate for disease spreading.
Here, we consider a simplicial contagion (describing the adoption of a behavior) that uni-directionally drives a simple contagion (describing a disease propagation). 
We show, both analytically and numerically, that, above a critical driving strength, such a driven simple contagion can exhibit both discontinuous transitions and bi-stability, absent otherwise.
Our results provide a novel route for a simple contagion process to display the phenomenology of a higher-order contagion, through a driving mechanism that may be hidden or unobservable in practical instances. 
\end{abstract}

%%%%%%%%%%%%%%%%%%%%%%%%%%%%%%%%%%%%%%%%%%%%%%%%%%%%%%%%%%%%%%%%	
\maketitle	
%%%%%%%%%%%%%%%%%%%%%%%%%%%%%%%%%%%%%%%%%%%%%%%%%%%%%%%%%%%%%%%%
%INTRODUCTION
%%%%%%%%%%%%%%%%%%%%%%%%%%%%%%%%%%%%%%%%%%%%%%%%%%%%%%%%%%%%%%%%

\section{Introduction} \label{sec:introduction}

Contagion processes have been widely studied using complex networks as the underlying structure supporting the propagation of diseases, innovation, and opinions~\cite{daley1964epidemics, pastor2001epidemic, rogers2003diffusion, barrat2008dynamical}.
The most studied examples include {\it simple contagion} models (where a contagion event can be caused by a single contact), such as the paradigmatic Susceptible-Infectious-Susceptible (SIS), widely used to describe the diffusion of a single pathogen in a population~\cite{vespignani2012modelling, pastor2015epidemic}.

In reality, however, contagion processes often co-exist and affect each other~\cite{sanz2014dynamics}. Infectious diseases can indeed display complex comorbidity interactions, in which the presence of a pathogen impacts the individual susceptibility towards another~\cite{wang2019coevolution}, like HIV increasing susceptibility to other sexually transmitted diseases~\cite{rottingen2001systematic}.
Modeling efforts in this direction include both cooperation~\cite{cai2015avalanche,chen2017fundamental,cui2017mutually} and competition~\cite{karrer2011competing,poletto2015characterising, li2022competing} between diseases.     
However, to date, models of interacting contagion processes have been developed under two main assumptions: 
({\it i}) the processes are simple contagions, and ({\it ii}) their interaction is symmetric, that is, bi-directional and of equal strength. 
Within these restrictions, cooperative models can display a discontinuous transition to the epidemic state~\cite{chen2017fundamental}, and become indistinguishable at the mean-field level from {\it complex contagion} models describing social reinforcement~\cite{hebert2020macroscopic} (where exposure to multiple sources presenting the same stimulus is needed for the contagion to occur
\cite{centola2007complex}).

Interactions between spreading processes are naturally not restricted to infectious diseases: a social behavior can also dramatically impact the spread of a disease~\cite{funk2010modelling,perra2011towards,granell2013dynamical, scarpino2016effect,perra2021non}. 
A current and cogent example is the impact of the adoption of \revision{risky} behaviors
(\revision{no} hand washing, \revision{no} masks, \revision{no} self-isolation \revision{or reduction of face-to-face contacts}) during the COVID pandemic~\cite{perra2021non}. 
Motivated by this example, we challenge both restrictions described above. 
First, it is known that reinforcement mechanisms influence social behavior so that models of simple contagion---that assume independent pairwise exposures---do not offer the most adequate description~\cite{centola2007complex}. 
\textit{Simplicial contagion} has been proposed as an alternative approach to account for simultaneous exposures via group-contagion events~\cite{iacopini2019simplicial, battiston2020networks}. 
Such group (``higher-order") contributions induce discontinuous transitions, bi-stability and critical mass phenomena even for single processes~\cite{barrat2022social, landry2020effect, ferrazdearruda2021phase, matamalas2020abrupt, st2022influential}.
Second, most contagion processes do not interact in a symmetric way. 
This can happen, e.g., for diseases with very different time scales~\cite{ventura2021role}, or when considering interactions between a disease and the adoption of prudent behaviors \cite{scarpino2016effect}, which is instead driven by a phenomenologically and analytically different social contagion process.\newline 

Here, we show that a simple contagion (describing infectious disease spreading) can exhibit the characteristics of a simplicial contagion when it is cooperatively driven by a simplicial contagion (describing the spread of a \revision{risky} social behavior). 
Namely, a simple contagion in the epidemic-free regime can exhibit an abrupt transition to the epidemic regime, as well as bi-stability, if the cooperative driving by the social process is stronger than a critical value. 
In particular, in the asymmetrically driven case, discontinuous transitions can only take place when the driving process is simplicial, contrary to the case of symmetric interactions.
We describe the phase diagram of the system through a mean-field (MF) approach, complemented by \revision{the numerical integration of coupled Markov-chain equations}, and provide an analytical expression for the critical value of the cooperation. Finally, we identify effective infectivities as markers of the abrupt driven transition by rewriting the MF equations as a simple contagion with effective parameters.

\section{Results} \label{sec:results}

%%%%%%%%%%%%%%%%%%%%%%%%%%%%%%%%%%%%%%%%%%%%%%%%%%%%%%%%%%%%%%%%
%THE MODEL
%%%%%%%%%%%%%%%%%%%%%%%%%%%%%%%%%%%%%%%%%%%%%%%%%%%%%%%%%%%%%%%%
\subsection{Model for interacting simplicial contagion processes} \label{sec:model}

\begin{figure}[t]
	\centering
	\includegraphics[width=\linewidth]{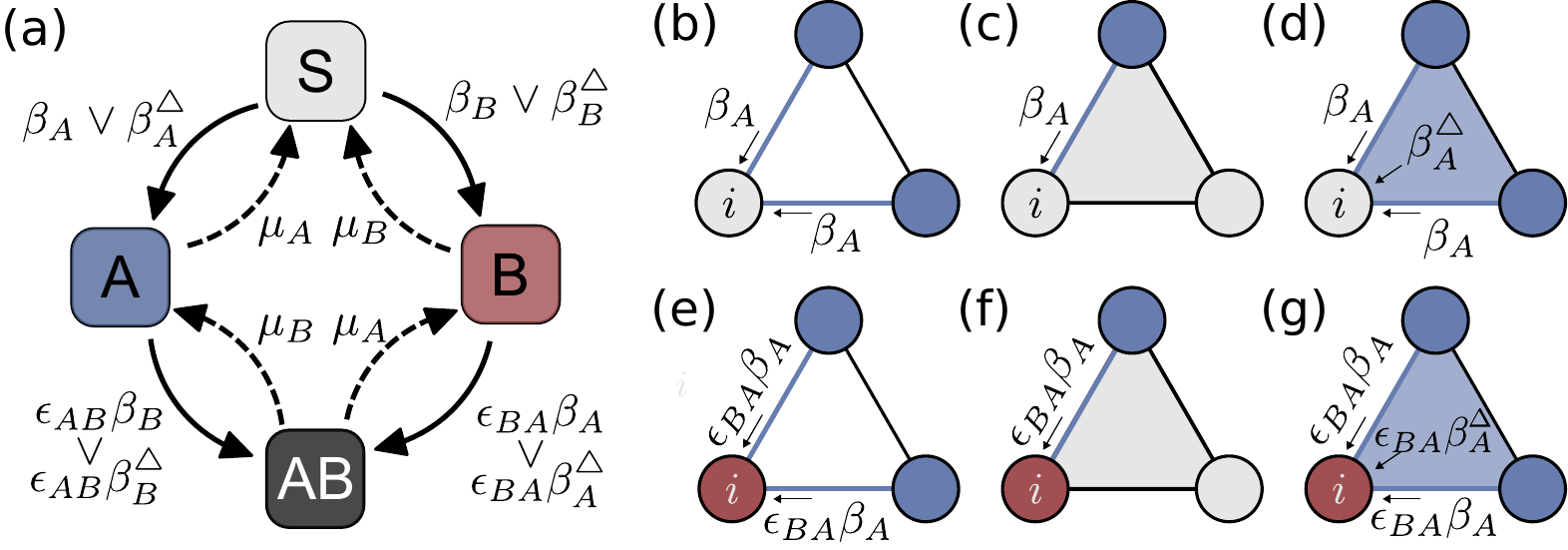}
	\vspace{-1em}
	\caption{\textbf{The model of interacting simplicial contagions.} (a) Transition probabilities between the compartments: susceptible ($S$, \textit{gray}), infected exclusively by one disease ($A$ or $B$, respectively \textit{blue/red}) or by both ($AB$, \textit{black}). (b)-(d) A susceptible node $i$ can acquire $A$ after a contact with an infectious \revision{$k$-hyperedge} (this also includes $AB$ individuals). In (d), since $i$ is part of a 2-simplex composed by two other infectious nodes, the infection can come both from each of the two \revision{1-hyperedges} (links) with probability $\beta_A$ and from the \revision{2-hyperedge} with probability $\beta_A^{\triangle}$. (e)-(g) If $i$ is already infected with $B$, the probability of getting $A$ for each contact is affected by the coupling factor $\epsBA$. The same rules symmetrically apply to $B$ instead of $A$.}
	\label{fig:simplicialinteractionssummary}
	\vspace{-1em}
\end{figure}

We consider a model for two interacting spreading processes, denoted as $A$ and $B$, which also include simplicial contagions~\cite{iacopini2019simplicial, barrat2022social}. 
Individuals are represented by a set of $N$ nodes 
%$\{ n_i \}_{i=1}^N$ 
that can each be in one of four compartments, following the standard SIS framework~\cite{pastor2015epidemic}: those susceptible to both diseases ($S$), infected exclusively by one of the two diseases (either $A$ or $B$), or by both ($AB$) [see Fig.~\ref{fig:simplicialinteractionssummary}(a)]. 
The compartment membership of each node $i$ is encoded in three binary variables $x_i^{\gamma} \in \{0, 1\}$, where $\gamma \in \{A, B, AB\}$. 
If node $i$ is in state $\gamma$ then $x_i^{\gamma}=1$, otherwise it is zero: each node has either one non-zero or all zero variables. 
The density of nodes in state $\gamma$ is given, at each time $t$, by $\rho_{\gamma} (t) = \frac1N \sum_{i=1}^N x_i^{\gamma} (t)$. 
The densities $\vec{\rho}(t) = \{ \rho_A(t), \rho_B(t), \rho_{AB}(t) \}$ serve as macroscopic order parameters (with $\rho_S(t) = 1 - \rho_A(t) - \rho_B(t) - \rho_{AB}(t)$ the density of susceptible individuals). 

Nodes can interact in pairs or larger groups, so that contagion events, which cause nodes to change compartment, take place on top of a contact structure that allows for higher-order (non-pairwise) interactions~\cite{lambiotte2019networks, battiston2020networks, bick2021higher, battiston2021physics}. 
We mathematically represent a group encounter as a $k$-hyperedge, a set of $k+1$ interacting nodes~\cite{torres2021and}.
For simplicity, we allow for interactions up to dimension $k=2$, i.e., on 1-hyperedges (links) and 2-hyperedges (triangles). 
Six parameters---three for each disease---yield contagion and recovery probabilities (Fig.~\ref{fig:simplicialinteractionssummary}).
The infectivity of disease $x \in \{A, B\}$ at order $k=1$, $\beta_{x,1} \equiv \beta_x$, is the probability per unit time for a node $i$ susceptible to pathogen $x$ to acquire $x$ from an ``infectious'' $1$-hyperedge it is part of [Fig.~\ref{fig:simplicialinteractionssummary}(b)-(d)]. Similarly, $\beta_{x,2} \equiv \beta_x^{\triangle}$ control infections coming from $2$-hyperedges [Fig.~\ref{fig:simplicialinteractionssummary}(d)]. Note that all other nodes in the hyperedge need to be infectious for the hyperedge to be considered so. 
Finally, $\mu_x\in[0,1]$ denotes the standard spontaneous recovery probability (from $x$) per unit time.

The interaction between the two contagion processes is controlled via two additional non-negative parameters, \revision{the coupling factors} $\epsAB$ and $\epsBA$ that multiply the transition probabilities to a double infection ($AB$) from a single infection ($A$ or $B$). For example, the transition $B\rightarrow AB$ occurs with probability $\epsBA \beta_A$ from a pairwise contact with $A$ [see Fig.~\ref{fig:simplicialinteractionssummary}(e)-(g)].  
The two processes cooperate if $\epsilon_{xx^{\prime}} > 1$ and compete if $\epsilon_{xx^{\prime}}<1$, while they are independent if  $\epsilon_{xx^{\prime}} = 1$. Note that the symmetry $\epsAB = \epsBA$ does not need to hold. 
Furthermore, although the model is defined on a generic higher-order structure, we focus here on simplicial complexes, a particular class of hypergraphs~\cite{battiston2020networks}. 
In a simplicial complex $\cal K$, by definition, groups of nodes are called simplices and respect downward closure: each sub-simplex $\nu \subset\sigma$ built from subsets of a simplex $\sigma\subset\cal K$ is also part of the complex $\cal K$ [in an infectious 2-simplex thus, contagion can occur both through the 1-hyperedges contained and
through the 2-hyperedge itself, see Fig.~\ref{fig:simplicialinteractionssummary}(d),(g)]. 
We make this choice for coherence with previous work \cite{iacopini2019simplicial}, but it can be relaxed to more general hypergraphs \cite{landry2020effect, barrat2022social, ferrazdearruda2021phase, st2022influential} without affecting the MF results. 
%

%%%%%%%%%%%%%%%%%%%%%%%%%%%%%%%%%%%%%%%%%%%%%%%%%%%%%%%%%%%%%%%%
%MEAN-FIELD APPROACH
%%%%%%%%%%%%%%%%%%%%%%%%%%%%%%%%%%%%%%%%%%%%%%%%%%%%%%%%%%%%%%%%
\subsection{Mean-field description} \label{sec:meanfield}

We consider the MF description of the model, obtained under a homogeneous mixing hypothesis~\cite{kiss2017mathematics}. 
\revision{For simplicity, we assume identical recovery rates for the two processes, that is $\mu_A = \mu_B = \mu$. In fact, for $\mu_A  \neq  \mu_B$ the equations can be simply refactored in term of a new parameter $\delta=\mu_A/\mu_B$ leaving the asymptotic dynamics unchanged (see Sup. Mat.~I).} We also introduce the rescaled infectivity parameters $\lambda_x = \beta_x \left< k \right> / \mu$ and  $\lambda_x^{\triangle} = \beta_x^{\triangle} \left< k_{\triangle} \right> / \mu$, for $x \in \{A, B\}$, where $\left< k \right>$ and $\left< k_{\triangle} \right>$ respectively denote the average numbers of 1- and 2-hyperedges incident on a node. \revision{After rescaling time by $\mu$, the general mean-field equations describing the evolution of the densities are:
\begin{subequations}\label{eq:MF}
	\begin{align}
\dot \rho_A =& - \rho_A + \lambda_A \rho_S (\rho_A + \rho_{AB}) 
+ \lambda_A^{\triangle} \rho_S (\rho_A + \rho_{AB})^2 \nonumber \\
&+ \rho_{AB} - \epsAB \lambda_B \rho_A (\rho_B + \rho_{AB})  \nonumber \\
&- \epsAB \lambda_B^{\triangle} \rho_A (\rho_B + \rho_{AB})^2 \\
\dot \rho_B =& - \rho_B + \lambda_B \rho_S (\rho_B + \rho_{AB})
+ \lambda_B^{\triangle} \rho_S (\rho_B + \rho_{AB})^2  \nonumber \\
&+ \rho_{AB} - \epsBA \lambda_A \rho_B (\rho_A + \rho_{AB}) \nonumber \\
&- \epsBA \lambda_A^{\triangle} \rho_B (\rho_A + \rho_{AB})^2  \\
\dot \rho_{AB} =& -2 \rho_{AB}+ \epsAB \lambda_B \rho_A (\rho_B + \rho_{AB}) \nonumber \\
&+ \epsAB \lambda_B^{\triangle} \rho_A (\rho_B + \rho_{AB})^2 \nonumber \\
&+ \epsBA \lambda_A \rho_B (\rho_A + \rho_{AB}) \nonumber\\ 
&+ \epsBA \lambda_A^{\triangle} \rho_B (\rho_A + \rho_{AB})^2      
\end{align}
\end{subequations}
with the additional condition that 
\begin{equation}
\rho_S = 1 - \rho_A - \rho_B - \rho_{AB} . \label{eq:tot}
\end{equation}}

%%%%%%%%%%%%%%%%%%%%%%%%%%%%%%%%%%%%%%%%%%%%%%%%%%%%%%%%%%%%%%%%
%SIMPLE CONTAGION LOOKS COMPLEX VIA COOPERATION
%%%%%%%%%%%%%%%%%%%%%%%%%%%%%%%%%%%%%%%%%%%%%%%%%%%%%%%%%%%%%%%%
\revision{In the following,} we focus on a simplicial contagion $A$ \revision{(representing a risky social behavior)} that cooperatively and uni-directionally drives a simple contagion $B$ \revision{(representing a disease)}. We thus set $\lambda_A^{\triangle}>0$, $\lambda_B^{\triangle}=0$, $\epsAB > 1$ and $\epsBA=1$. 

\revision{In this scenario, it is convenient to consider the total density of infectious individuals for each contagion, regardless of whether they are also infected by the other one. Formally, we introduce two new variables $\rho_{A_{\text{tot}}} = \rho_A + \rho_{AB}$ and $\rho_{B_{\text{tot}}} = \rho_B + \rho_{AB}$. In other words, $\rho_{A_{\text{tot}}}$ is the total density of people with a risky behaviour, having been infected by $B$ ($\rho_{AB}$), or not ($\rho_{A}$). Similarly, $\rho_{B_{\text{tot}}}$ is the total density of people infected by disease $B$, having a risky behavior ($\rho_{AB}$), or not ($\rho_{B}$).} After introducing these two variables, we end up with the \revision{system of} coupled equations \revision{(see Appendix \ref{app:MF})}:
\begin{subequations} \label{eq:MF_tot}
\begin{align}
\dot \rho_{A_{\text{tot}}} &= \rho_{A_{\text{tot}}} \left[ -1 + \lambda_A (1 - \rho_{A_{\text{tot}}}) \right. \nonumber \\
& + \lambda_A^{\triangle} \rho_{A_{\text{tot}}}  (1 - \rho_{A_{\text{tot}}}) ] \label{eq:MF_tot_Atot} , \\
\dot \rho_{B_{\text{tot}}} &= \rho_{B_{\text{tot}}} \left[ - 1 + \lambda_B (1 - \rho_{B_{\text{tot}}}) \right.  \nonumber \\
 & \left. + \lambda_B (\epsAB - 1 ) (\rho_{A_{\text{tot}}} -  \rho_{AB})  \right] , \label{eq:MF_tot_Btot} \\
\dot \rho_{AB} &= -2 \rho_{AB}+ \epsAB \lambda_B (\rho_{A_{\text{tot}}} - \rho_{AB}) \rho_{B_{\text{tot}}} \nonumber \\
&  +  \lambda_A (\rho_{B_{\text{tot}}} - \rho_{AB}) \rho_{A_{\text{tot}}} +  \lambda_A^{\triangle} (\rho_{B_{\text{tot}}} - \rho_{AB}) \rho_{A_{\text{tot}}}^2  \label{eq:MF_tot_AB}  .
\end{align} 
\end{subequations}
Equations~\eqref{eq:MF_tot} include two known specific cases. 
First, without any interaction between the processes ($\epsAB = \epsBA = 1$), $\rho_{B_{\text{tot}}}$ and $\rho_{A_{\text{tot}}}$ evolve independently as a simple and simplicial contagion~\cite{iacopini2019simplicial}, respectively. 
Second, by considering only pairwise interactions, $\lambda_A^{\triangle}= 0 = \lambda_B^{\triangle}$, $A$ and $B$ evolve as interacting simple contagions~\cite{wang2019coevolution}. 
In the general case we consider ($\epsBA = 1$), the dynamics of $\rho_{A_{\text{tot}}}$ is decoupled from the other two variables and drives them.

\begin{figure}[t]
	\centering
	\includegraphics[width=0.99\linewidth]{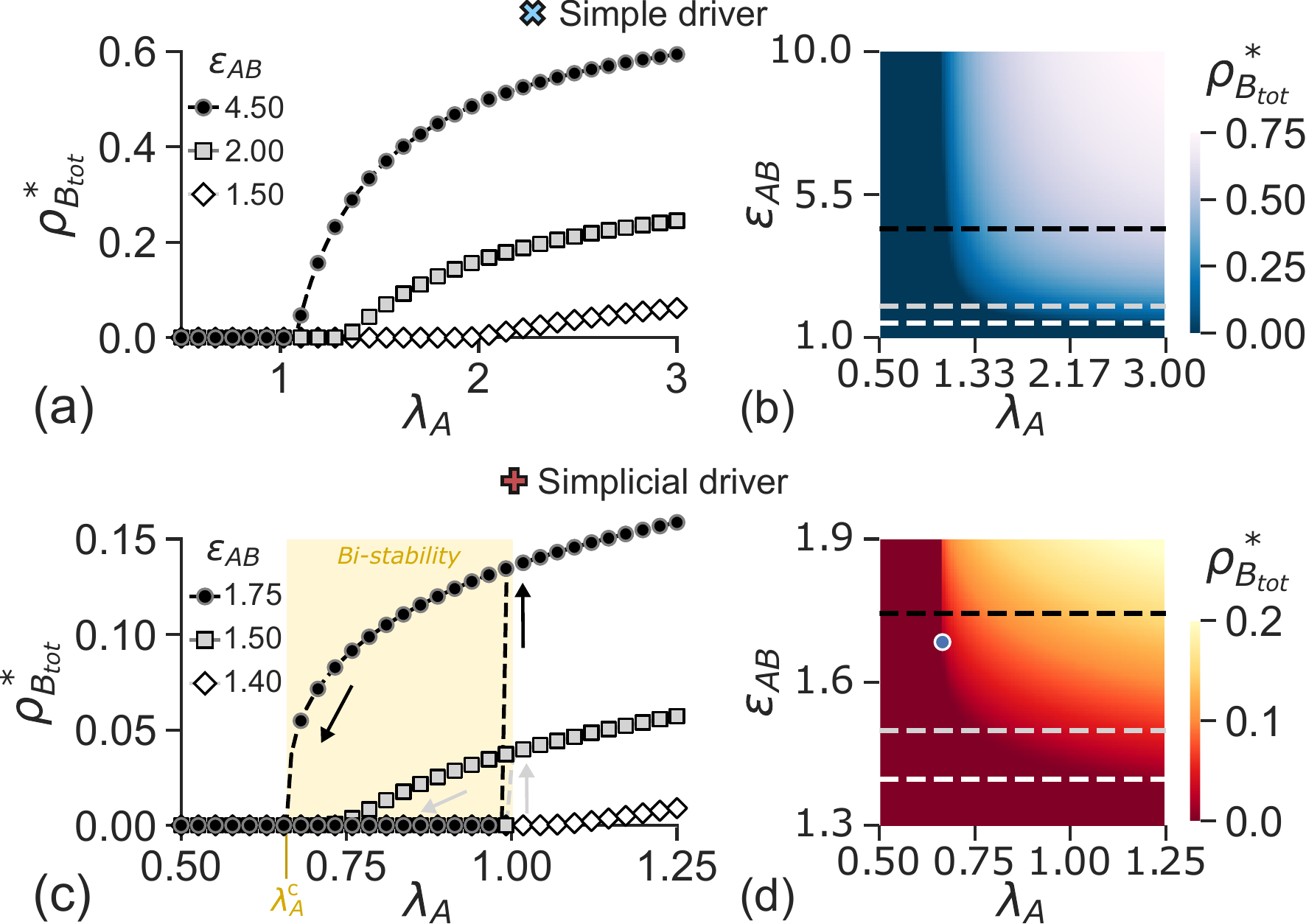}
	%\vspace{-1em}
	\caption{\textbf{Abrupt transition induced by a simplicial driver.} A simplicial driver process for $A$, \revision{with $\lambda_A^\triangle=2.5$}, can induce a discontinuous transition (c,d), contrary to a simple driver (a,b), \revision{with $\lambda_A^\triangle=0$} [$\lambda_B=0.8$, $\lambda_B^{\triangle}=0$].
	Note the different scales on the horizontal axes. (a,c)
	Stationary solutions $\rho^{*}_{B_{\text{tot}}}$ \revision{of the MF Eqs.~\eqref{eq:MF_tot}} plotted as a function of the rescaled pairwise infectivity $\lambda_A$ for three values of the driving strength $\epsAB$. In (c), the transition of the simple contagion $B$ becomes discontinuous above a critical value of cooperation $\epsAB^c$. (b,d) Heatmaps of $\rho^{*}_{B_{\text{tot}}}$ as a function of $\lambda_A$ and $\epsAB$. Dashed horizontal lines correspond to the selected $\epsAB$ values shown in (a) and (c) respectively. The blue dot in (d) highlights the critical point $(\lambda_A^c, \epsAB^c)$. The blue and red crosses represent a visual hint to locate the results within the full phase diagram of Fig.~\ref{fig:eps_crit}.
	}
	\label{fig:explosivetransitionphasediagram}
	\vspace{-1em}
\end{figure}

We first study the non-equilibrium stationary state (NESS) reached by the system \revision{of Eqs.}~\eqref{eq:MF_tot} at large times $\rho^*_{B_{\text{tot}}} = \lim_{t \to \infty} \rho_{B_{\text{tot}}}(t)$, by numerical integration. Figure~\ref{fig:explosivetransitionphasediagram} shows the resulting $\rho^*_{B_{\text{tot}}}$ values and their transitions. 
The most interesting case is given by $\lambda_B < 1$
as, without a driving process $A$, the simple contagion process $B$ would be in the epidemic-free absorbing state ($\rho^*_{B_\text{tot}}=0$). 
We thus illustrate how the $B$ NESS
$\rho^*_{B_\text{tot}}$ depends on the parameters of the driver $A$, on the coupling $\epsAB$, and how it can transition to the epidemic active state, despite $\lambda_B < 1$.

For $\lambda_{A}^{\triangle} \le 1$, we always obtain a continuous transition for $\rho^*_{B_{\text{tot}}}$ [see Fig.~\ref{fig:explosivetransitionphasediagram}(a,b)]. 
On the other hand, if $\lambda_{A}^{\triangle} > 1$, the driven process can exhibit a discontinuous transition [see Fig.~\ref{fig:explosivetransitionphasediagram}(c,d)]. 
More precisely, the transition changes from continuous to discontinuous when the coupling parameter $\epsAB$ becomes larger than a critical value $\epsAB^c$. 
Above this threshold [$\epsAB > \epsAB^c$, black circles in Fig.~\ref{fig:explosivetransitionphasediagram}(c)], there is a discontinuous transition at a critical value $\lambda_A^c$ that does not depend on $\epsAB$. 
For weaker cooperation [$\epsAB < \epsAB^c$, white and gray symbols in Fig.~\ref{fig:explosivetransitionphasediagram}(c)], a continuous transition occurs from $\rho_{B_{\text{tot}}}^* = 0$ to the epidemic state $\rho_{B_{\text{tot}}}^*>0$ when $\lambda_A$ crosses another critical value $\lambda_A^{'c} \ge \lambda_A^c$. \revision{In other words, $\lambda_A^{'c}$ is the critical value of  $\lambda_A$ at which the continuous transition occurs, for $\epsAB < \epsAB^c$, and its value decreases as $\epsAB$ increases---e.g. from $\lambda_A^{'c} \approx 1.1$ (white symbols) to $\lambda_A^{'c} \approx 0.7$ (gray symbols). On the contrary, $\lambda_A^c$ (yellow label) is where the discontinuous transition occurs, for $\epsAB > \epsAB^c$, and does not depend on $\epsAB$. 
%In fact, it only depends on $A$'s parameters $\lambda_{A}$ and $\lambda_{A}^{\triangle}$ as we shall show below. 
Note that $\lambda_A^{'c} \to \lambda_A^c$ in the limit $\epsAB \to \epsAB^c$.}

\revision{It is important to note that} the epidemic-free absorbing state $\rho_{B_{\text{tot}}}^{*}=0$ remains stable as long as $\lambda_A < 1$. \revision{As a consequence, there is a region of bi-stability $\lambda_A^c < \lambda_A < 1$ (region shaded with yellow background) for $\epsAB > \epsAB^c$. Bi-stability can also be observed when $\epsAB < \epsAB^c$ in the region $\lambda_A^{'c} < \lambda_A < 1$ (gray symbols) as long as $\lambda_A^{'c} < 1$. For $\lambda_A^{'c} \ge 1$ (white symbols), the continuous transition occurs for values larger than one and there is no region of bi-stability. Finally, when $\lambda_A^{'c} < 1$, the stability of the absorbing state implies the existence of a forward discontinuous transition at $\lambda_A^{'c} = 1$ (upward arrows).}
\revision{In conclusion}, the simple contagion $B$ exhibits characteristics of a simplicial contagion---an abrupt transition and bi-stability---due to the driving of the simplicial contagion $A$. 

To analytically explain this behavior, we need to find the NESS by setting $\dot \rho_{x} = 0$ \revision{(see Appendix \ref{app:MF_fixed_points} for additional details).}
Solving $\dot \rho_{A_{\text{tot}}} = 0$, as Eq.~\eqref{eq:MF_tot_Atot} exactly maps back to the single simplicial contagion analyzed in Ref.~\cite{iacopini2019simplicial}, leads to a trivial solution $\rho^*_{A_{\text{tot}}}=0$ and two other NESS $\rho_{A_{\text{tot}}}^{*, \pm}$. Similarly, solving the full two-dimensional system $(\rho_{B_{\text{tot}}}, \rho_{AB})$ leads to the absorbing state $(0,0)$ and the implicit solutions for $\rho^*_{B_{\text{tot}}}$:
\begin{equation}
\rho_{B_{\text{tot}}}^{*, \pm} = 1 - \frac{1}{\lambda_B} + (\rho_{A_{\text{tot}}}^{*, \pm} - \rho_{AB}^{*, \pm})  (\epsAB - 1 ).
\label{eq:fixed_point_btot}
\end{equation}
Equation~\eqref{eq:fixed_point_btot} implicitly contains two solutions $\pm$ from $\rho_{A_{\text{tot}}}^{*, \pm}$ and $\rho_{AB}^{*, \pm}$. 

\begin{figure}[t]
	\centering
	\includegraphics[width=0.9\linewidth]{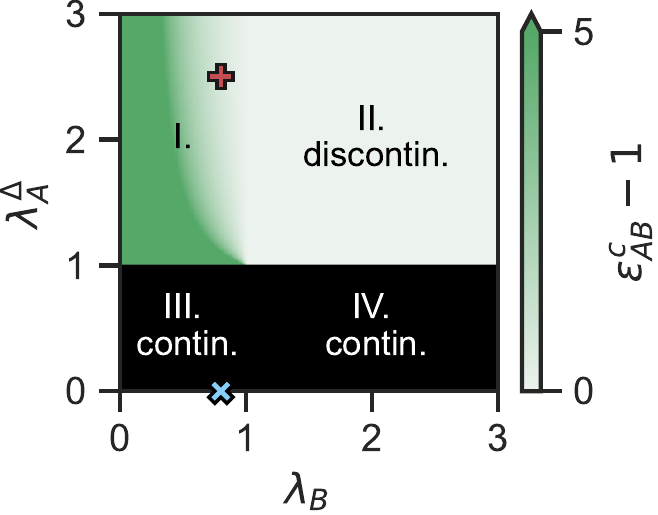}
	\vspace{-1em}
	\caption{\textbf{Phase diagram of the system.} The $(\lambda_B, \lambda_{A}^{\triangle})$ parameter space exhibits four regions. In region I ($\lambda_B \le 1$, $\lambda_{A}^{\triangle} > 1$), $\rho^*_{B_{\text{tot}}}$ undergoes an abrupt transition if the driving cooperation is strong enough, $\epsAB > \epsAB^c$. The value of $\epsAB^c$ is represented by shades of green. For visual clarity, the green scale is truncated at a maximum value of 5, so that larger values are represented by the same color as 5. The red cross corresponds to the case shown in Figs.~\ref{fig:explosivetransitionphasediagram}(c,d). In region II ($\lambda_B > 1$, $\lambda_{A}^{\triangle} > 1$), $\epsAB^c=1$ and the transition is discontinuous for all $\epsAB > 1$. For $\lambda_{A}^{\triangle} \le 1$, that is regions III and IV, the transition is always continuous. The blue cross indicates the case shown in Figs.~\ref{fig:explosivetransitionphasediagram}(a,b).}
	\label{fig:eps_crit}
	\vspace{-1em}
\end{figure}

\revision{Using the implicit solutions Eqs. (\ref{eq:fixed_point_btot}), we can understand the behavior shown in Fig.~\ref{fig:explosivetransitionphasediagram}. 
If $\lambda_{A}^{\triangle} < 1$ [Figs.~\ref{fig:explosivetransitionphasediagram}(a,b)], $\rho^*_{A_{\text{tot}}}$ exhibits a continuous transition at $\lambda_A=1$, below which it is zero (see \cite{iacopini2019simplicial}). This implies that $\rho_{AB}$ also goes to the absorbing state if $\lambda_A<1$: if nobody is infected by $A$, nobody can be infected by both $A$ and $B$. As a consequence, the term coming from the cooperation vanishes: $\rho_{A_{\text{tot}}}^{*, \pm} - \rho_{AB}^{*, \pm} = 0$. 
Hence, from Eq. \eqref{eq:fixed_point_btot}, we have $\lim_{\lambda_A \to 1^-} \rho_{B_{\text{tot}}}^{*, \pm} = 1 - \frac{1}{\lambda_B}  \le 0$ (recall $\lambda_B < 1$). If negative, it is not a valid solution and only the absorbing state is. As $\lambda_A$ increases above $1$, $\rho_{A_{\text{tot}}}^{*, +}$ increases continuously~\cite{iacopini2019simplicial}, leading $\rho_{B_{\text{tot}}}^{*, \pm}$ to also cross continuously $0$ at a certain $\lambda_A \ge 1$.
For $\lambda_{A}^{\triangle} \ge 1$ instead, $\rho^*_{A_{\text{tot}}}$ has a discontinuous transition at $\lambda_A = \lambda_A^c$ which implies that $\rho_{AB}^*$ has one too. Consequently, from Eq. \eqref{eq:fixed_point_btot}, since $\lambda_B < 1$, we have $\lim_{\lambda_A \to \lambda_A^{c,-}} \rho_{B_{\text{tot}}}^{*} = 0$, but $\lim_{\lambda_A \to \lambda_A^{c,+}} \rho_{B_{\text{tot}}}^{*, \pm} > 0$ above a certain value $\epsAB > \epsAB^c$. Hence, this critical value $\epsAB^c$ can be derived analytically by solving $\rho_{B_{\text{tot}}}^{*,+} = 0$ at $\lambda_A^c$. In other words, we find the critical driving strength $\epsAB^c$ by finding the curve $\rho_{B_{\text{tot}}}^{*,+}$, between the gray and black curves in Fig. \ref{fig:explosivetransitionphasediagram}(c), such that it reaches zero at $\lambda_A^c$. 
This corresponds to the case $\lambda_B \le 1$, $\lambda_{A}^{\triangle} > 1$ which we denoted region I in Fig. \ref{fig:eps_crit}. 
Similarly, if $\lambda_B > 1$ instead (with $\lambda_{A}^{\triangle} > 1$, region II), it suffices solving $\rho_{B_{\text{tot}}}^{*,+} = 1 - 1/\lambda_B$ at $\lambda_A^c$ because $1 - 1/\lambda_B$ is now the pre-transition NESS. In summary, the discontinuity is controlled by the term coming from the cooperation, $(\rho_{A_{\text{tot}}}^{*, \pm} - \rho_{AB}^{*, \pm})  (\epsAB - 1 )$ which will be discontinuous if $A$ has a discontinuous transition. }

In short, the nature of the transition depends on the new second term induced by the driving \revision{in Eq. \eqref{eq:fixed_point_btot}} and, in particular, on the $\lambda_A$ value at which it becomes positive.
This fact, allow us to obtain the critical $\epsAB^c$ at which the discontinuous transition in $B$ becomes possible (Fig.~\ref{fig:eps_crit}):
\begin{equation}
\epsAB^c = \left\{ 
\begin{array}{cl}
     \frac{\sqrt{\lambda_{A}^{\triangle}} - \lambda_B}{\left( \sqrt{\lambda_{A}^{\triangle}} - 1 \right) \lambda_B} & \text{in region I ($\lambda_B \le 1$, 
     $\lambda_{A}^{\triangle} >1$), } \\
     1 & \text{in region II ($\lambda_B > 1$, 
     $\lambda_{A}^{\triangle} >1$).  }
\end{array}
\right.
\end{equation}
In region I, increasing $\lambda_{A}^{\triangle}$ or $\lambda_B$ makes $\epsAB^c$ decrease, so that discontinuous transitions are obtained for smaller values of the driving strength $\epsAB$. 
In fact, $\epsAB^c \to +\infty$ as $\lambda_{A}^{\triangle} \to 1$ or $\lambda_B \to 0$. In region II, all values of cooperation $\epsAB > 1$ yield a discontinuous transition.
Finally, no critical value of cooperation can be defined in regions III and IV ($\lambda_{A}^{\triangle} \le 1$), where transitions are always continuous.

%%%%%%%%%%%%%%%%%%%%%%%%%%%%%%%%%%%%%%%%%%%%%%%%%%%%%%%%%%%%%%%%
%EFFECTIVE FORMALISM 
%%%%%%%%%%%%%%%%%%%%%%%%%%%%%%%%%%%%%%%%%%%%%%%%%%%%%%%%%%%%%%%%
\subsection{Effective formalism} \label{sec:effective}
\revision{In this section, we devise an effective contagion theory to highlight the origins of the observed transitions. In particular, we saw that,} for $\epsAB>\epsAB^c$---that is, when simplicial behavior is possible for $B$---the driven process $B$ will exhibit discontinuous transitions as a function of $\lambda_A$. 
We illustrate how this phenomenology emerges by rewriting the dynamics of $B$ as an effective simple contagion following Ref.~\cite{hebert2020macroscopic}. 

\revision{First, note that we can rewrite} the MF equation of the single simplicial $\rho_{A_{\text{tot}}}$ from Eq.~\eqref{eq:MF_tot_Atot} as a simple contagion 
\revision{%
\begin{equation}
\dot  \rho_{A_{\text{tot}}} = - \rho_{A_{\text{tot}}} + \tilde \lambda_A \, \rho_{A_{\text{tot}}} [1 - \rho_{A_{\text{tot}}}] ,
\label{eq:A_effective}
\end{equation}%
}
with effective infectivity $\tilde \lambda_A = \lambda_A + \lambda_A^{\triangle} \rho_{A_{\text{tot}}}$. \revision{As expected, the effective infectivity depends on both the simple and simplicial infectivities. Since Eq. \eqref{eq:A_effective} is written as a simple contagion, its well known stationary solutions given by $1 - 1 / \tilde \lambda_A$ and the effective infectivity also has a critical value of 1.}

Similarly, we can also rewrite Eq.~\eqref{eq:MF_tot_Btot} of the driven $\rho_{B_{\text{tot}}}$  as a simple contagion
\begin{equation}
\dot  \rho_{B_{\text{tot}}} = - \rho_{B_{\text{tot}}} + \tilde \lambda_B \, \rho_{B_{\text{tot}}} [1 - \rho_{B_{\text{tot}}}] ,
\label{eq:B_effective}
\end{equation}
with effective infectivity
\begin{equation}
\tilde \lambda_B = \lambda_B + \lambda_B (\epsAB - 1) \frac{1}{1 - \rho_{B_{\text{tot}}}} \rho_A .
\end{equation}
\revision{Since we observed characteristics of the driver $A$ in the driven contagion $B$, we may want to further cast its effective infectivity into a form similar to that of $A$:  $\tilde \lambda_B = \lambda_B + \lambda_B^{\triangle} \rho_{B_{\text{tot}}}$. } 
% %
\revision{This is achieved} by defining an effective simplicial infectivity
\begin{equation}
\tilde \lambda_B^{\triangle} = \lambda_B (\epsAB - 1) \frac{\rho_A}{\rho_{B_{\text{tot}}} (1 - \rho_{B_{\text{tot}}})}   ,
\end{equation}
which implicitly depends on $\lambda_A$ and $\lambda_A^{\triangle}$ through $\rho_A$.

\begin{figure}[t]
    \centering
    \includegraphics[width=\linewidth]{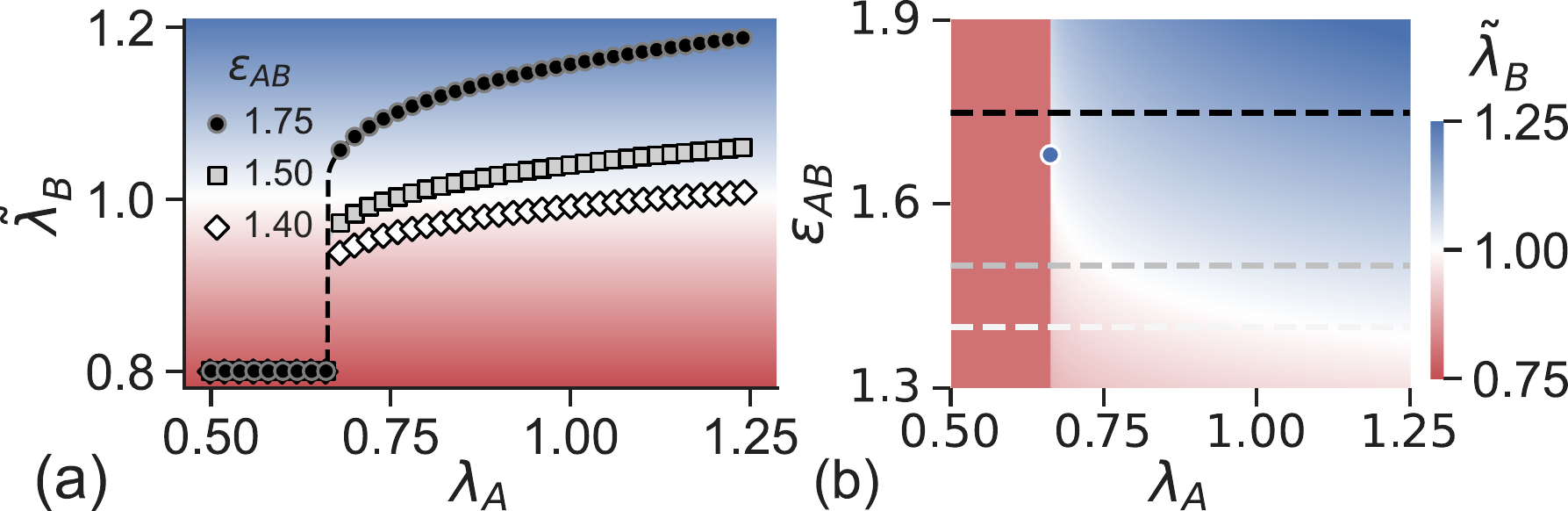}
    \vspace{-1em}
    \caption{\textbf{The discontinuous nature of the driven contagion $B$ can be determined from its effective infectivity $\tilde \lambda_B$.} \revision{(a) We show $\tilde \lambda_B$ against the infectivity $\lambda_A$ for several values of cooperation $\epsAB$}, corresponding to the curves in Fig.~\ref{fig:explosivetransitionphasediagram}(c). \revision{The full phase diagram as a function of both $\lambda_A$ and $\epsAB$ is shown as a heatmap in panel (b), where the dot corresponds} to the critical point $(\lambda_A^c, \epsAB^c)$. \revision{The background color in (a) corresponds to the colorbar in (b). It corresponds to the values of the vertical axis and highlight visually the critical value $\tilde \lambda_B = 1$} }
    \vspace{-1em}
    \label{fig:effective}
\end{figure}

If there is no interaction ($\epsAB = 1$), we recover $\tilde \lambda_B = \lambda_B$ \revision{because} the effective simplicial infectivity vanishes, $\tilde \lambda_B^{\triangle} = 0$, as expected. 
\revision{More importantly, since $\rho_{B_{\text{tot}}}$ evolves according to the effectively simple contagion of Eq. \eqref{eq:B_effective}, its stationary solution is given by $1 - 1/\tilde \lambda_B$ which yields a critical value $\tilde \lambda_B = 1$. This} can help distinguish between the transitions observed in Fig.~\ref{fig:explosivetransitionphasediagram}. 
Indeed,
\revision{below the critical value, $\tilde \lambda_B < 1$, the only stable NESS of $\rho_{B_{\text{tot}}}$ is $0$, and above it there is also a positive solution.}
Thus, the driven contagion $B$ has a transition to an epidemic state if and only if $\tilde \lambda_B$ crosses 1 as $\lambda_A$ increases. 
\revision{Finally and most importantly}, this transition is discontinuous if and only if the transition of $\tilde \lambda_B$ across the values one is discontinuous. 
This can be seen by comparing the three curves in Fig.~\ref{fig:effective}(a)\revision{---crossing the value one (white background) discontinuously (black) or continuously (gray and white)---}with the corresponding curves for $\rho_{B_{\text{tot}}}^*$ in Fig.~\ref{fig:explosivetransitionphasediagram}(c).

%%%%%%%%%%%%%%%%%%%%%%%%%%%%%%%%%%%%%%%%%%%%%%%%%%%%%%%%%%%%%%%%
%TEMPORAL DYNAMICS
%%%%%%%%%%%%%%%%%%%%%%%%%%%%%%%%%%%%%%%%%%%%%%%%%%%%%%%%%%%%%%%%
\subsection{Temporal properties} \label{sec:temporal}

So far, we lack information about the temporal trajectories, which, in practical settings, are often the only data available. 
Consider observing the spread of $B$ via $\rho_{B_{\text{tot}}}(t)$, while the driving social contagion process $A$ remains unobservable. 
Interestingly, the observed $B$ evolves differently depending on the initial conditions of the hidden process $A$.
%\revision{---we here discuss the bi-stability aspect of the model}.

We show the phenomenology described beyond the homogeneous mixing hypothesis, by shifting to a Markov-chain formalism~\cite{gomez2010discrete, soriano2019markovian}.
\revision{With this microscopic approach we can  encode any interaction structure between nodes---contrary to MF approaches that assume homogeneous mixing of the population---while keeping the computational cost lower than the one required for Monte Carlo simulations. The complete Markov-chain description of our model can be found in Appendix \ref{app:markov}. 
We build a synthetic random simplicial complex up to dimension 2 by means of the generative model introduced in Ref.~\cite{iacopini2019simplicial}. This model, a direct extension of Erd\"os-R\'enyi-like models for graphs, allows to generate a simplicial complex starting from a number of nodes that get randomly connected to form simplices. 
%While the creation of simplices at different orders is controlled via a probability vector, it is also possible to reverse engineer the problem by choosing the desired generalised degrees instead. 
We here generate a simplicial complex with $N=2000$ nodes having $\langle k\rangle=20$ and $\langle k^{\triangle}\rangle=6$ and integrate the associated Markov equations to follow the temporal evolution of the system.} 

\begin{figure}[t]
	\centering
	\includegraphics[width=0.99\linewidth]{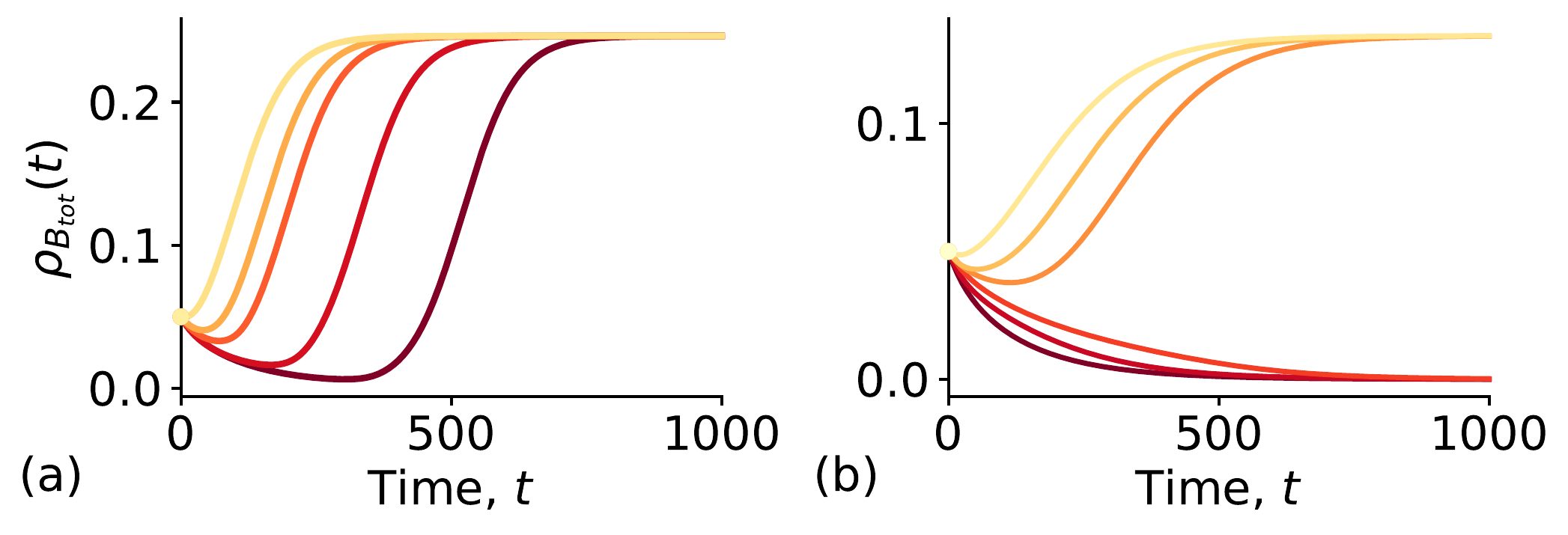}
	\vspace{-1em}
	\caption{
	\textbf{The temporal evolution of the simple contagion $B$ is affected by the initial conditions of the (hidden) simplicial driver $A$.} We show $\rho_{B_{\text{tot}}}$ over time, \revision{resulting from the numerical integration of the Markov-chain equations for a simplicial complex with $N=2000$ nodes, $\langle k\rangle=20$ and $\langle k^{\triangle}\rangle=6$}, for $A$ in (a) the endemic region, $\lambda_A=1.2$, and (b) the bi-stable region, $\lambda_A=0.7$. Shades of red \revision{from dark to light represent a range of initial conditions of the driver $\rho_A(0)$ from $0.001$ to $0.35$ [see Sup. Mat. Fig. S1 for the temporal evolution of $\rho_A(t)$].} In (b), the simple contagion process $B$ can reach one of two stationary states, depending on the initial conditions of the driver $A$. Other parameters are set to $\lambda_B=0.8$, $\lambda_B^\triangle=0$, $\epsAB=1.75$, and $\lambda_A^\triangle=2.5$. }
	\label{fig:temporal_dynamics}
	\vspace{-1em}
\end{figure}

We consider the scenario of the black curve ($\epsAB=1.75$) of Fig.~\ref{fig:explosivetransitionphasediagram}(b), that is, with a simplicial driver $A$ ($\lambda_A^\triangle=2.5$). 
We fix all other parameters, including the initial condition $\rho_{B_{\text{tot}}}(0)$, but vary the initial condition of the driver, $\rho_{A_{\text{tot}}}(0)$. 
As shown in Fig.~\ref{fig:temporal_dynamics},
if the driver contagion $A$ is in the endemic regime but not in the bi-stability region (e.g., $\lambda_A=1.2$), $\rho_{B_{\text{tot}}}$ reaches the same NESS for all values of $\rho_{A_{\text{tot}}}(0)$, but with different transient dynamics and even non-monotonic evolutions [Fig.~\ref{fig:temporal_dynamics}(a)].  Moreover, if the simplicial driver is in the bi-stability region ($\lambda_A=0.7$), it induces bi-stability in $B$: $\rho_{B_{\text{tot}}}(t)$ can reach two different states, depending on the driving initial condition, even though all ``visible'' $B$ parameters are fixed [Fig.~\ref{fig:temporal_dynamics}(b)]. 
Note that this bi-stability emerges only if the driving is simplicial with $\lambda_A^\triangle > 1$ (see also Sup. Mat. Fig.~S1). 

%%%%%%%%%%%%%%%%%%%%%%%%%%%%%%%%%%%%%%%%%%%%%%%%%%%%%%%%%%%%%%%%
% CONCLUSIONS
%%%%%%%%%%%%%%%%%%%%%%%%%%%%%%%%%%%%%%%%%%%%%%%%%%%%%%%%%%%%%%%%
\section{Discussion} \label{sec:discussion}

In conclusion, our results highlight that an abrupt transition in the observed process can occur as a function of the control parameter of a second---potentially hidden---driver process.
Consider an observer of an epidemic process of unknown nature. A natural intervention would try to reduce the intrinsic infectivity of the spreading pathogen, e.g. through pharmaceutical interventions or reduction of social contacts (sanitary lockdowns): this would however lead only to a continuous change in the incidence. 
However, if the spread is driven by an underlying complex contagion, then acting on the hidden driver process (e.g. \revision{trying to reduce} the social adoption of \revision{risky} behaviors) could more effectively lead to an abrupt transition to the epidemic-free state (if the interaction is strong enough $\epsAB > \epsAB^c$). 
\revision{Finally, different populations could be characterised by different properties of the hidden behavioral contagion process (different values of $\lambda_A$ and $\lambda_A^\Delta$ values), thus leading to a large diversity of temporal evolutions, and---potentially---of  final outcomes of the pathogen's spread, without the need for different intrinsic infectivity properties of the pathogen across these populations.}

\revision{Results also suggest that other driving spreading processes could yield a similar phenomenology if they exhibit a discontinuous transition (e.g. \cite{gomez2016explosive}), inducing a change from a continuous to a discontinuous transition in the driven process. 
We note in this context that the framework of Ref.~\cite{kuehn2021universal} suggests a universal route to abrupt transitions, achieved through the addition of a control parameter to a process that displays a continuous phase transition. 
However, the situation that we have explored here broadens the picture.
Indeed, if both spreading processes are simple contagions, it appears that a bi-directional interaction (leading to a feedback loop) is an additional necessary condition for a discontinuous transition to emerge. 
In the case of a uni-directional coupling, instead, the driving process needs to be itself simplicial with bi-stability.}
Our results thus provide a different route to the emergence of abrupt transitions \revision{in epidemic-like processes} due to the asymmetric coupling of the contagion dynamics, as opposed to the addition of a control parameter~\cite{kuehn2021universal}. \revision{This resonates with recent results in synchronization phenomena \cite{khanra2021explosive}.}
The exact conditions under which these routes apply to coupled systems in general would be an interesting direction for future work.
\revision{Another interesting perspective would consist in the analysis of real-world data and the development of tools to detect the footprints of simple, complex, or coupled processes from observed time series \cite{brett2020detecting} in order to discriminate them or, potentially, perform full reconstruction~\cite{peixoto2019network}.}

%%%%%%%%%%%%%%%%%%%%%%%%%%%%%%%%%%%%%%%%%%%%%%%%%%%%%%%%%%%%%%%%
% ACKNOWLEDGEMENTS
%%%%%%%%%%%%%%%%%%%%%%%%%%%%%%%%%%%%%%%%%%%%%%%%%%%%%%%%%%%%%%%%
\section*{Acknowledgements}
I.I. acknowledges support from the James S. McDonnell Foundation $21^{\text{st}}$ Century Science Initiative Understanding Dynamic and Multi-scale Systems - Postdoctoral Fellowship Award. A.B. acknowledges support from the Agence Nationale de la Recherche (ANR) project DATAREDUX (ANR-19-CE46-0008).  M.L. and G.P. acknowledge partial support from the Intesa Sanpaolo Innovation Center during the preparation of this work.
\textbf{Code availability.} \revision{The code used in this study is available at \url{https://github.com/iaciac/interacting-simplagions}.}

%%%%%%%%%%%%%%%%%%%%%%%%%%%%%%%%%%%%%%%%%%%%%%%%%%%%%%%%%%%%%%%%
% APPENDIX
%%%%%%%%%%%%%%%%%%%%%%%%%%%%%%%%%%%%%%%%%%%%%%%%%%%%%%%%%%%%%%%%
\begin{appendix}
\section{Derivation of MF description} \label{app:MF}

As explained in the main text, we focus on the case $\epsBA=1$, $\lambda_B^{\triangle}=0$, so that Eqs.~\eqref{eq:MF} becomes
\begin{subequations}
	\begin{align}
\dot \rho_A =& - \rho_A + \lambda_A \rho_S (\rho_A + \rho_{AB})
+ \lambda_A^{\triangle} \rho_S (\rho_A + \rho_{AB})^2 \nonumber \\
&+ \rho_{AB} - \epsAB \lambda_B \rho_A (\rho_B + \rho_{AB}) , \\
\dot \rho_B =& - \rho_B + \lambda_B \rho_S (\rho_B + \rho_{AB}) + \rho_{AB} \nonumber \\
&- \lambda_A \rho_B (\rho_A + \rho_{AB}) 
- \lambda_A^{\triangle} \rho_B (\rho_A + \rho_{AB})^2 , \\
\dot \rho_{AB} =& -2 \rho_{AB}+ \epsAB \lambda_B \rho_A (\rho_B + \rho_{AB}) \nonumber \\
&+ \lambda_A \rho_B (\rho_A + \rho_{AB}) %\nonumber\\ 
+ \lambda_A^{\triangle} \rho_B (\rho_A + \rho_{AB})^2     . 
\end{align}
\label{SI:eq:MF_driven}
\end{subequations}
Then, we apply the following change of variables: $\rho_{A_{\text{tot}}} = \rho_A + \rho_{AB}$, $\rho_{B_{\text{tot}}} = \rho_B + \rho_{AB}$. This yields 
\begin{subequations}
\begin{align}
\dot \rho_{A_{\text{tot}}} ={}& (- \rho_A - \rho_{AB}) + \lambda_A \rho_{A_{\text{tot}}} [1 - \rho_{B_{\text{tot}}} - \rho_A + \rho_B] \nonumber \\ 
&+ \lambda_A^{\triangle} \rho_{A_{\text{tot}}}^2  [1 - \rho_{B_{\text{tot}}} - \rho_A + \rho_B]  , \\
\dot  \rho_{B_{\text{tot}}} ={}&  (- \rho_B - \rho_{AB}) + \lambda_B \rho_{B_{\text{tot}}} [1 - \rho_{A_{\text{tot}}} - \rho_B + \epsAB \rho_A]  , \\	
\dot \rho_{AB} ={}& -2 \rho_{AB}+ \epsAB \lambda_B \rho_A \rho_{B_{\text{tot}}} +  \lambda_A \rho_B \rho_{A_{\text{tot}}} +  \lambda_A^{\triangle} \rho_B \rho_{A_{\text{tot}}}^2 .
\end{align}
\end{subequations}
We further rewrite this by replacing all remaining $\rho_A$ and $\rho_B$, and using the identity $1 - \rho_{B_{\text{tot}}} - \rho_A + \rho_B = 1 - \rho_{A_{\text{tot}}}$, 
\begin{subequations}
\begin{align}
\dot \rho_{A_{\text{tot}}} ={}& - \rho_{A_{\text{tot}}} + \lambda_A \rho_{A_{\text{tot}}} [1 - \rho_{A_{\text{tot}}}]  + \lambda_A^{\triangle} \rho_{A_{\text{tot}}}^2  [1 - \rho_{A_{\text{tot}}}] , \\
\dot  \rho_{B_{\text{tot}}} ={}&  - \rho_{B_{\text{tot}}} + \lambda_B \rho_{B_{\text{tot}}} [1 - \rho_{A_{\text{tot}}} - \rho_{B_{\text{tot}}} +  \rho_{AB} \nonumber \\
&+ \epsAB (\rho_{A_{\text{tot}}} - \rho_{AB})] , \\
\dot \rho_{AB} ={}& -2 \rho_{AB}+ \epsAB \lambda_B (\rho_{A_{\text{tot}}} - \rho_{AB}) \rho_{B_{\text{tot}}} \nonumber \\
&+  \lambda_A (\rho_{B_{\text{tot}}} - \rho_{AB}) \rho_{A_{\text{tot}}} +  \lambda_A^{\triangle} (\rho_{B_{\text{tot}}} - \rho_{AB}) \rho_{A_{\text{tot}}}^2 ,
\end{align}
\end{subequations}
which can be refactored to obtain Eqs.~\eqref{eq:MF_tot} from the main text.

\section{Derivation of the MF fixed points} \label{app:MF_fixed_points}
Equation \eqref{eq:MF_tot_Atot} is the same as the simplicial contagion from Ref.~\cite{iacopini2019simplicial}, and its non-trivial solutions are
\begin{equation} 
\rho_{A_{\text{tot}}}^{*, \pm}  = \frac{ (\lambda_A^{\triangle} - \lambda_A) \pm \sqrt{ (\lambda_A - \lambda_A^{\triangle})^2 + 4  \lambda_A^{\triangle}  (\lambda_A - 1)} }{2 \lambda_A^{\triangle}  } . 
\label{SI:eq:simplagion}
\end{equation}
For this isolated case, we know that $\lambda_A^{\triangle}$ controls the type of transition to the epidemic state \cite{iacopini2019simplicial}. 
That is, for $\lambda_A^{\triangle} \le 1$, the bifurcation diagram has a continuous transition at $\lambda_A=1$ from $\rho^*_{A_{\text{tot}}}=0$ to the epidemic state $\rho_{A_{\text{tot}}}^{*, +}$. 
When instead $\lambda_A^{\triangle}>1$, a discontinuous transition to $\rho_{A_{\text{tot}}}^{*, +}$ occurs at  $\lambda_A^c = -\lambda_A^{\triangle} + 2 \sqrt{\lambda_A^{\triangle}} \le 1$. 
The epidemic-free state remains stable for $\lambda_A \le 1$, but becomes unstable above: 
This leads to bi-stability in the parameter region $\{\lambda_A^{\triangle}>1, \lambda_A^c \le \lambda_A \le 1\}$. 
The discontinuous transition is therefore the direct consequence of a sufficiently strong three-body (higher-order) interaction in $A$ ($\lambda_A^{\triangle}>1$).

The remaining two-dimensional system $(\rho_{B_{\text{tot}}}, \rho_{AB})$ can be solved analytically by hand or with the help of software such as {\it Mathematica}~\cite{Mathematica}. As discussed, the implicit solution for Eq.~\eqref{eq:MF_tot_Btot} is given by Eq.~\eqref{eq:fixed_point_btot}. 

To solve for $\rho_{AB}$, we rewrite Eq.~\eqref{eq:MF_tot_AB} by factorizing and setting the left-hand side to zero:
\begin{align}
0 =& -2 \rho_{AB} + \epsAB \lambda_B (\rho_{A_{\text{tot}}} - \rho_{AB}) \rho_{B_{\text{tot}}} \nonumber \\ 
&+  \lambda_A (\rho_{B_{\text{tot}}} - \rho_{AB}) \rho_{A_{\text{tot}}} +  \lambda_A^{\triangle} (\rho_{B_{\text{tot}}} - \rho_{AB}) \rho_{A_{\text{tot}}}^2 , \\
=& \rho_{AB} \left[ -2 - \epsAB \lambda_B \rho_{B_{\text{tot}}} - \lambda_A \rho_{A_{\text{tot}}} - \lambda_A^{\triangle} \rho_{A_{\text{tot}}}^2 \right] \\ \nonumber
+& \rho_{A_{\text{tot}}} \rho_{B_{\text{tot}}} \left[ \epsAB \lambda_B + \lambda_A + \lambda_A^{\triangle} \rho_{A_{\text{tot}}} \right] , 
\end{align} 
from which we already see that $\rho_{AB}^*=0$ if $\rho_{A_{\text{tot}}} = 0$ or $\rho_{B_{\text{tot}}} = 0$. Now, we inject the expression of $\rho_{B_{\text{tot}}}^*$ from Eq.~\eqref{eq:fixed_point_btot} and cast the equation into quadratic form in $\rho_{AB}$:
\begin{equation}
0 = A \rho_{AB}^2 + B \rho_{AB} + C ,
\end{equation}
where 
\begin{align}
A ={}& + \epsAB \lambda_B E^-_{AB} , \\
B ={}& - 2 - \epsAB \lambda_B (\Lambda^-_B + E^-_{AB} \rho_{A_{\text{tot}}}^*) \\ \nonumber
&- ( \lambda_A + E^-_{AB} K) \rho_{A_{\text{tot}}}^* - \lambda_A^{\triangle} \rho_{A_{\text{tot}}}^{* 2}  , \\
C ={}& \rho_{A_{\text{tot}}}^* K  (\Lambda^-_B + E^-_{AB} \rho_{A_{\text{tot}}}^*) .
\end{align}
To shorten the notation, we have also defined
\begin{align}
E^-_{AB} ={}& \epsAB - 1 , \\
\Lambda^-_i ={}& 1 - 1 / \lambda_i , \\
K ={}& \epsAB \lambda_B + \lambda_A + \lambda_A^{\triangle} \rho_{A_{\text{tot}}}^* .
\end{align}
The non-zero solutions for $\rho_{AB}$ is the standard quadratic solution 
\begin{equation}
\rho_{AB}^{*, \pm} = \frac{-B \pm \sqrt{B^2 - 4AC}}{2A} ,
\end{equation}
which, unfolded, is an expression in terms of the parameters of the system only. These together with $\rho_{A_{\text{tot}}}^*$ can be reinjected into Eq.~\eqref{eq:fixed_point_btot} for $\rho_{B_{\text{tot}}}^*$ to close the system.

\section{Markov-chain approach} \label{app:markov}

Here, we write a system of coupled Markov-chain equations which govern the microscopic evolution of our model~\cite{gomez2010discrete, soriano2019markovian}. More precisely, we can write down the conditional probability $P(x_i^\gamma(t+1)=1|{\bf x}(t), {\bf\theta}, {\bf A})\equiv p^i_\gamma(t)$ of finding each node $i$ in state $\gamma=\{S,A,B,AB\}$ at time $t+1$ given the probability vector representing the status of all nodes at time $t$ ${\bf x}(t)={x_i^\gamma(t)}$, the model parameters ${\bf\theta}=\{\beta_A,\beta_A^\triangle,\beta_B,\beta_B^\triangle,\mu_A,\mu_B,\epsAB,\epsBA \}$, and the structure ${\bf A}$. Using the simplified notation $p^i_\gamma(t)$, we impose that, at each time,
\begin{equation}
p_S^i(t)=1-p_A^i(t)-p_B^i(t)-p_{AB}^i(t).
\end{equation}
The Markov-chain equations for the three states are the following:

\begin{subequations}\label{SI:eq:MMCA}
	\begin{align}
	p_{AB}^i(t+1) = &+\revision{p_B}^i(t)(1-\mu_B)(1-q_A^i(t)) \nonumber \\
	&+ p_A^i(t)(1-\mu_A)(1-q_B^i(t)) \nonumber \\
	&+ p_{AB}^i(t)(1-\mu_A)(1-\mu_B) ,\\
	p_A^i(t+1) = &+p_{AB}^i(t)\mu_B(1-\mu_A) \nonumber \\
	&+p_A^i(t)(1-\mu_A)q_B^i(t) \nonumber \\
	&+p_B^i(t)\mu_B(1-q_A^i(t)) \nonumber \\
	&+p_S^i(t)(1-q_{AB}^i(t))f_A^i(t) ,\\
	p_B^i(t+1) = &+p_{AB}^i(t)\mu_A(1-\mu_B) \nonumber \\
	&+p_B^i(t)(1-\mu_B)q_A^i(t) \nonumber \\
	&+p_A^i(t)\mu_A(1-q_B^i(t)) \nonumber \\
	&+p_S^i(t)(1-q_{AB}^i(t))f_B^i(t) .
	\end{align}
\end{subequations}

The different $q_x^i(t)$ denote the probability of node $i$ not being infected by disease $x$ by any of the simplices it participates in. Considering again only contributions up to $D=2$, we have:

\begin{widetext}
\begin{subequations}\label{SI:eq:MMCA_q}
\begin{align}
    q_A^i(t) =& \prod_{j\in\mathcal{V}} \Big\{ 1-a_{ij}\epsBA\beta_A[p_A^j(t)+p_{AB}^j(t)] \Big\} \prod_{j,l\in\mathcal{V}}\Bigl[1 -a_{ijl}\epsBA\beta_A^{\triangle}[p_A^j(t)+p_{AB}^j(t)] [p_A^l(t)+p_{AB}^l(t)] \Bigr] ,\\
    q_B^i(t) =& \prod_{j\in\mathcal{V}} \Big\{ 1-a_{ij}\epsAB\beta_B[p_B^j(t)+p_{AB}^j(t)] \Big\}  \prod_{j,l\in\mathcal{V}}\Bigl[1 -a_{ijl}\epsAB\beta_B^{\triangle}[p_B^j(t)+p_{AB}^j(t)] [p_B^l(t)+p_{AB}^l(t)] \Bigr] ,\\
    q_{AB}^i(t)
    =& \prod_{j\in\mathcal{V}} \Big\{ 1-a_{ij}\Bigl[\beta_A p_A^j(t) + \beta_B p_B^j(t) +[\beta_A(1-\beta_B)+\beta_B(1-\beta_A)+\beta_A\beta_B] p_{AB}^j(t)\Bigr]\Big\} \nonumber \\
    &\prod_{j,l\in\mathcal{V}}\Big\{1-a_{ijl}
    \Bigl[\beta_A^{\triangle}[p_A^j(t)p_A^l(t) +p_A^j(t)p_{AB}^l(t)+p_{AB}^j(t)p_A^l(t)] +\beta_B^{\triangle}[p_B^j(t)p_B^l(t)+p_B^j(t)p_{AB}^l(t)+p_{AB}^j(t)p_B^l(t)] \nonumber \\
    &+[\beta_A^{\triangle}(1-\beta_B^{\triangle})+\beta_B^{\triangle}(1-\beta_A^{\triangle})+\beta_A^{\triangle}\beta_B^{\triangle}p_{AB}^j(t)p_{AB}^l(t)]\Bigr]\Big\} \nonumber \\
    =& \prod_{j\in\mathcal{V}} \Big\{ 1-a_{ij}\Bigl[\revision{\beta_A}[p_A^j(t)+p_{AB}^j(t)] +\beta_B[p_B^j(t)+p_{AB}^j(t)]-\revision{\beta_A\beta_B[p_{AB}^j(t)]}\Bigr]\Big\} \nonumber \\
    &\prod_{j,l\in\mathcal{V}}\Big\{1-a_{ijl}\Bigl[\beta_A^{\triangle}[p_A^j(t)+p_{AB}^j(t)][p_A^l(t)+p_{AB}^l(t)] +\beta_B^{\triangle}[p_B^j(t)+p_{AB}^j(t)][p_B^l(t)+p_{AB}^l(t)] \revision{-\beta_A^{\triangle}\beta_B^{\triangle}p_{AB}^j(t)p_{AB}^l(t)}\Bigr]\Big\},
\end{align}
\end{subequations}
\end{widetext}
where the first product of each equation accounts for the contagion through the links of the simplicial complex $\cal K$. These links are fully specified by means of the standard adjacency matrix $\{a_{ij}\}$, whose elements $a_{ij}=0,1$ denote the absence or presence of a link $(i,j)$. Similarly, the second product accounts for the contagion of $i$ through the 2-simplices of $\cal K$ (triangles), which are analogously specified by the elements of the adjacency tensor $\{a_{ijl}\}$. This tensor is the 3-dimensional version of the adjacency matrix, in which a non-zero element denotes the presence of a 2-simplex $(i,j,l)$.

Finally, the factors $f_A^i(t)$ and $f_B^i(t)$ in Eq.~\eqref{SI:eq:MMCA} denote the probability of transitioning from state $S$ to one of the states $A$ or $B$ when exposed simultaneously to both pathogens. Assuming an equal probability for both diseases~\cite{soriano2019markovian}, we can write:

\begin{subequations}\label{eq:fa_fb}
\begin{align}
    f_A^i(t)=\frac{\bar{q}_A^i(t)(1-0.5\bar{q}_B^i(t))}{\bar{q}_A^i(t)(1-\revision{0.5}\bar{q}_B^i(t))+\bar{q}_B^i(t)(1-0.5\bar{q}_A^i(t))}\\
    f_B^i(t)=\frac{\bar{q}_B^i(t)(1-0.5\bar{q}_A^i(t))}{\bar{q}_A^i(t)(1-\revision{0.5}\bar{q}_B^i(t))+\bar{q}_B^i(t)(1-0.5\bar{q}_A^i(t))}
\end{align}
\end{subequations}
where $\bar{q}_A^i(t)$ and $\bar{q}_B^i(t)$ correspond to $1-q_A^i(t)$ and $1-q_B^i(t)$, as given by Eqs.~\eqref{SI:eq:MMCA_q}, after setting $\epsAB=\epsBA=1$.

\end{appendix}

%%%%%%%%%%%%%%%%%%%%%%%%%%%%%%%%%%%%%%%%%%%%%%%%%%%%%%%%%%%%%%%%
% BIBLIOGRAPHY
%%%%%%%%%%%%%%%%%%%%%%%%%%%%%%%%%%%%%%%%%%%%%%%%%%%%%%%%%%%%%%%%
%\bibliography{bib}

\begin{thebibliography}{43}%
	\makeatletter
	\providecommand \@ifxundefined [1]{%
		\@ifx{#1\undefined}
	}%
	\providecommand \@ifnum [1]{%
		\ifnum #1\expandafter \@firstoftwo
		\else \expandafter \@secondoftwo
		\fi
	}%
	\providecommand \@ifx [1]{%
		\ifx #1\expandafter \@firstoftwo
		\else \expandafter \@secondoftwo
		\fi
	}%
	\providecommand \natexlab [1]{#1}%
	\providecommand \enquote  [1]{``#1''}%
	\providecommand \bibnamefont  [1]{#1}%
	\providecommand \bibfnamefont [1]{#1}%
	\providecommand \citenamefont [1]{#1}%
	\providecommand \href@noop [0]{\@secondoftwo}%
	\providecommand \href [0]{\begingroup \@sanitize@url \@href}%
	\providecommand \@href[1]{\@@startlink{#1}\@@href}%
	\providecommand \@@href[1]{\endgroup#1\@@endlink}%
	\providecommand \@sanitize@url [0]{\catcode `\\12\catcode `\$12\catcode
		`\&12\catcode `\#12\catcode `\^12\catcode `\_12\catcode `\%12\relax}%
	\providecommand \@@startlink[1]{}%
	\providecommand \@@endlink[0]{}%
	\providecommand \url  [0]{\begingroup\@sanitize@url \@url }%
	\providecommand \@url [1]{\endgroup\@href {#1}{\urlprefix }}%
	\providecommand \urlprefix  [0]{URL }%
	\providecommand \Eprint [0]{\href }%
	\providecommand \doibase [0]{https://doi.org/}%
	\providecommand \selectlanguage [0]{\@gobble}%
	\providecommand \bibinfo  [0]{\@secondoftwo}%
	\providecommand \bibfield  [0]{\@secondoftwo}%
	\providecommand \translation [1]{[#1]}%
	\providecommand \BibitemOpen [0]{}%
	\providecommand \bibitemStop [0]{}%
	\providecommand \bibitemNoStop [0]{.\EOS\space}%
	\providecommand \EOS [0]{\spacefactor3000\relax}%
	\providecommand \BibitemShut  [1]{\csname bibitem#1\endcsname}%
	\let\auto@bib@innerbib\@empty
	%</preamble>
	\bibitem [{\citenamefont {Daley}\ and\ \citenamefont
		{Kendall}(1964)}]{daley1964epidemics}%
	\BibitemOpen
	\bibfield  {author} {\bibinfo {author} {\bibfnamefont {D.~J.}\ \bibnamefont
			{Daley}}\ and\ \bibinfo {author} {\bibfnamefont {D.~G.}\ \bibnamefont
			{Kendall}},\ }\bibfield  {title} {\bibinfo {title} {Epidemics and rumours},\
	}\href {https://doi.org/https://doi.org/10.1038/2041118a0} {\bibfield
		{journal} {\bibinfo  {journal} {Nature}\ }\textbf {\bibinfo {volume} {204}},\
		\bibinfo {pages} {1118} (\bibinfo {year} {1964})}\BibitemShut {NoStop}%
	\bibitem [{\citenamefont {Pastor-Satorras}\ and\ \citenamefont
		{Vespignani}(2001)}]{pastor2001epidemic}%
	\BibitemOpen
	\bibfield  {author} {\bibinfo {author} {\bibfnamefont {R.}~\bibnamefont
			{Pastor-Satorras}}\ and\ \bibinfo {author} {\bibfnamefont {A.}~\bibnamefont
			{Vespignani}},\ }\bibfield  {title} {\bibinfo {title} {Epidemic spreading in
			scale-free networks},\ }\href {https://doi.org/10.1103/PhysRevLett.86.3200}
	{\bibfield  {journal} {\bibinfo  {journal} {Phys. Rev. Lett.}\ }\textbf
		{\bibinfo {volume} {86}},\ \bibinfo {pages} {3200} (\bibinfo {year}
		{2001})}\BibitemShut {NoStop}%
	\bibitem [{\citenamefont {Rogers}(2003)}]{rogers2003diffusion}%
	\BibitemOpen
	\bibfield  {author} {\bibinfo {author} {\bibfnamefont {E.}~\bibnamefont
			{Rogers}},\ }\href {https://books.google.at/books?id=9U1K5LjUOwEC} {\emph
		{\bibinfo {title} {Diffusion of Innovations, 5th Edition}}}\ (\bibinfo
	{publisher} {Free Press},\ \bibinfo {year} {2003})\BibitemShut {NoStop}%
	\bibitem [{\citenamefont {Barrat}\ \emph {et~al.}(2008)\citenamefont {Barrat},
		\citenamefont {Barth{\'e}lemy},\ and\ \citenamefont
		{Vespignani}}]{barrat2008dynamical}%
	\BibitemOpen
	\bibfield  {author} {\bibinfo {author} {\bibfnamefont {A.}~\bibnamefont
			{Barrat}}, \bibinfo {author} {\bibfnamefont {M.}~\bibnamefont
			{Barth{\'e}lemy}},\ and\ \bibinfo {author} {\bibfnamefont {A.}~\bibnamefont
			{Vespignani}},\ }\href {https://books.google.at/books?id=TmgePn9uQD4C} {\emph
		{\bibinfo {title} {Dynamical Processes on Complex Networks}}}\ (\bibinfo
	{publisher} {Cambridge University Press},\ \bibinfo {year}
	{2008})\BibitemShut {NoStop}%
	\bibitem [{\citenamefont {Vespignani}(2012)}]{vespignani2012modelling}%
	\BibitemOpen
	\bibfield  {author} {\bibinfo {author} {\bibfnamefont {A.}~\bibnamefont
			{Vespignani}},\ }\bibfield  {title} {\bibinfo {title} {Modelling dynamical
			processes in complex socio-technical systems},\ }\href
	{https://doi.org/https://doi.org/10.1038/nphys2160} {\bibfield  {journal}
		{\bibinfo  {journal} {Nat. Phys.}\ }\textbf {\bibinfo {volume} {8}},\
		\bibinfo {pages} {32} (\bibinfo {year} {2012})}\BibitemShut {NoStop}%
	\bibitem [{\citenamefont {Pastor-Satorras}\ \emph {et~al.}(2015)\citenamefont
		{Pastor-Satorras}, \citenamefont {Castellano}, \citenamefont {Van~Mieghem},\
		and\ \citenamefont {Vespignani}}]{pastor2015epidemic}%
	\BibitemOpen
	\bibfield  {author} {\bibinfo {author} {\bibfnamefont {R.}~\bibnamefont
			{Pastor-Satorras}}, \bibinfo {author} {\bibfnamefont {C.}~\bibnamefont
			{Castellano}}, \bibinfo {author} {\bibfnamefont {P.}~\bibnamefont
			{Van~Mieghem}},\ and\ \bibinfo {author} {\bibfnamefont {A.}~\bibnamefont
			{Vespignani}},\ }\bibfield  {title} {\bibinfo {title} {Epidemic processes in
			complex networks},\ }\href {https://doi.org/10.1103/RevModPhys.87.925}
	{\bibfield  {journal} {\bibinfo  {journal} {Rev. Mod. Phys.}\ }\textbf
		{\bibinfo {volume} {87}},\ \bibinfo {pages} {925} (\bibinfo {year}
		{2015})}\BibitemShut {NoStop}%
	\bibitem [{\citenamefont {Sanz}\ \emph {et~al.}(2014)\citenamefont {Sanz},
		\citenamefont {Xia}, \citenamefont {Meloni},\ and\ \citenamefont
		{Moreno}}]{sanz2014dynamics}%
	\BibitemOpen
	\bibfield  {author} {\bibinfo {author} {\bibfnamefont {J.}~\bibnamefont
			{Sanz}}, \bibinfo {author} {\bibfnamefont {C.-Y.}\ \bibnamefont {Xia}},
		\bibinfo {author} {\bibfnamefont {S.}~\bibnamefont {Meloni}},\ and\ \bibinfo
		{author} {\bibfnamefont {Y.}~\bibnamefont {Moreno}},\ }\bibfield  {title}
	{\bibinfo {title} {Dynamics of interacting diseases},\ }\href
	{https://doi.org/10.1103/PhysRevX.4.041005} {\bibfield  {journal} {\bibinfo
			{journal} {Phys. Rev. X}\ }\textbf {\bibinfo {volume} {4}},\ \bibinfo {pages}
		{041005} (\bibinfo {year} {2014})}\BibitemShut {NoStop}%
	\bibitem [{\citenamefont {Wang}\ \emph {et~al.}(2019)\citenamefont {Wang},
		\citenamefont {Liu}, \citenamefont {Liang}, \citenamefont {Hu},\ and\
		\citenamefont {Zhou}}]{wang2019coevolution}%
	\BibitemOpen
	\bibfield  {author} {\bibinfo {author} {\bibfnamefont {W.}~\bibnamefont
			{Wang}}, \bibinfo {author} {\bibfnamefont {Q.-H.}\ \bibnamefont {Liu}},
		\bibinfo {author} {\bibfnamefont {J.}~\bibnamefont {Liang}}, \bibinfo
		{author} {\bibfnamefont {Y.}~\bibnamefont {Hu}},\ and\ \bibinfo {author}
		{\bibfnamefont {T.}~\bibnamefont {Zhou}},\ }\bibfield  {title} {\bibinfo
		{title} {Coevolution spreading in complex networks},\ }\href
	{https://doi.org/https://doi.org/10.1016/j.physrep.2019.07.001} {\bibfield
		{journal} {\bibinfo  {journal} {Phys. Rep.}\ }\textbf {\bibinfo {volume}
			{820}},\ \bibinfo {pages} {1} (\bibinfo {year} {2019})}\BibitemShut {NoStop}%
	\bibitem [{\citenamefont {R{\o}ttingen}\ \emph {et~al.}(2001)\citenamefont
		{R{\o}ttingen}, \citenamefont {Cameron},\ and\ \citenamefont
		{Garnett}}]{rottingen2001systematic}%
	\BibitemOpen
	\bibfield  {author} {\bibinfo {author} {\bibfnamefont {J.-A.}\ \bibnamefont
			{R{\o}ttingen}}, \bibinfo {author} {\bibfnamefont {D.~W.}\ \bibnamefont
			{Cameron}},\ and\ \bibinfo {author} {\bibfnamefont {G.~P.}\ \bibnamefont
			{Garnett}},\ }\bibfield  {title} {\bibinfo {title} {A systematic review of
			the epidemiologie interactions between classic sexually transmitted diseases
			and hiv: How much really is known?},\ }\href
	{http://www.jstor.org/stable/44965414} {\bibfield  {journal} {\bibinfo
			{journal} {Sex. Transm. Dis.}\ ,\ \bibinfo {pages} {579}} (\bibinfo {year}
		{2001})}\BibitemShut {NoStop}%
	\bibitem [{\citenamefont {Cai}\ \emph {et~al.}(2015)\citenamefont {Cai},
		\citenamefont {Chen}, \citenamefont {Ghanbarnejad},\ and\ \citenamefont
		{Grassberger}}]{cai2015avalanche}%
	\BibitemOpen
	\bibfield  {author} {\bibinfo {author} {\bibfnamefont {W.}~\bibnamefont
			{Cai}}, \bibinfo {author} {\bibfnamefont {L.}~\bibnamefont {Chen}}, \bibinfo
		{author} {\bibfnamefont {F.}~\bibnamefont {Ghanbarnejad}},\ and\ \bibinfo
		{author} {\bibfnamefont {P.}~\bibnamefont {Grassberger}},\ }\bibfield
	{title} {\bibinfo {title} {Avalanche outbreaks emerging in cooperative
			contagions},\ }\href {https://doi.org/https://doi.org/10.1038/nphys3457}
	{\bibfield  {journal} {\bibinfo  {journal} {Nat. Phys.}\ }\textbf {\bibinfo
			{volume} {11}},\ \bibinfo {pages} {936} (\bibinfo {year} {2015})}\BibitemShut
	{NoStop}%
	\bibitem [{\citenamefont {Chen}\ \emph {et~al.}(2017)\citenamefont {Chen},
		\citenamefont {Ghanbarnejad},\ and\ \citenamefont
		{Brockmann}}]{chen2017fundamental}%
	\BibitemOpen
	\bibfield  {author} {\bibinfo {author} {\bibfnamefont {L.}~\bibnamefont
			{Chen}}, \bibinfo {author} {\bibfnamefont {F.}~\bibnamefont {Ghanbarnejad}},\
		and\ \bibinfo {author} {\bibfnamefont {D.}~\bibnamefont {Brockmann}},\
	}\bibfield  {title} {\bibinfo {title} {Fundamental properties of cooperative
			contagion processes},\ }\href
	{https://doi.org/https://doi.org/10.1088/1367-2630/aa8bd2} {\bibfield
		{journal} {\bibinfo  {journal} {New J. Phys.}\ }\textbf {\bibinfo {volume}
			{19}},\ \bibinfo {pages} {103041} (\bibinfo {year} {2017})}\BibitemShut
	{NoStop}%
	\bibitem [{\citenamefont {Cui}\ \emph {et~al.}(2017)\citenamefont {Cui},
		\citenamefont {Colaiori}, \citenamefont {Castellano} \emph
		{et~al.}}]{cui2017mutually}%
	\BibitemOpen
	\bibfield  {author} {\bibinfo {author} {\bibfnamefont {P.-B.}\ \bibnamefont
			{Cui}}, \bibinfo {author} {\bibfnamefont {F.}~\bibnamefont {Colaiori}},
		\bibinfo {author} {\bibfnamefont {C.}~\bibnamefont {Castellano}}, \emph
		{et~al.},\ }\bibfield  {title} {\bibinfo {title} {Mutually cooperative
			epidemics on power-law networks},\ }\href
	{https://doi.org/10.1103/PhysRevE.96.022301} {\bibfield  {journal} {\bibinfo
			{journal} {Phys. Rev. E}\ }\textbf {\bibinfo {volume} {96}},\ \bibinfo
		{pages} {022301} (\bibinfo {year} {2017})}\BibitemShut {NoStop}%
	\bibitem [{\citenamefont {Karrer}\ and\ \citenamefont
		{Newman}(2011)}]{karrer2011competing}%
	\BibitemOpen
	\bibfield  {author} {\bibinfo {author} {\bibfnamefont {B.}~\bibnamefont
			{Karrer}}\ and\ \bibinfo {author} {\bibfnamefont {M.~E.}\ \bibnamefont
			{Newman}},\ }\bibfield  {title} {\bibinfo {title} {Competing epidemics on
			complex networks},\ }\href
	{https://doi.org/https://doi.org/10.1103/PhysRevE.84.036106} {\bibfield
		{journal} {\bibinfo  {journal} {Phys. Rev. E}\ }\textbf {\bibinfo {volume}
			{84}},\ \bibinfo {pages} {036106} (\bibinfo {year} {2011})}\BibitemShut
	{NoStop}%
	\bibitem [{\citenamefont {Poletto}\ \emph {et~al.}(2015)\citenamefont
		{Poletto}, \citenamefont {Meloni}, \citenamefont {Van~Metre}, \citenamefont
		{Colizza}, \citenamefont {Moreno},\ and\ \citenamefont
		{Vespignani}}]{poletto2015characterising}%
	\BibitemOpen
	\bibfield  {author} {\bibinfo {author} {\bibfnamefont {C.}~\bibnamefont
			{Poletto}}, \bibinfo {author} {\bibfnamefont {S.}~\bibnamefont {Meloni}},
		\bibinfo {author} {\bibfnamefont {A.}~\bibnamefont {Van~Metre}}, \bibinfo
		{author} {\bibfnamefont {V.}~\bibnamefont {Colizza}}, \bibinfo {author}
		{\bibfnamefont {Y.}~\bibnamefont {Moreno}},\ and\ \bibinfo {author}
		{\bibfnamefont {A.}~\bibnamefont {Vespignani}},\ }\bibfield  {title}
	{\bibinfo {title} {Characterising two-pathogen competition in spatially
			structured environments},\ }\href
	{https://doi.org/https://doi.org/10.1038/srep07895} {\bibfield  {journal}
		{\bibinfo  {journal} {Sci. Rep.}\ }\textbf {\bibinfo {volume} {5}},\ \bibinfo
		{pages} {1} (\bibinfo {year} {2015})}\BibitemShut {NoStop}%
	\bibitem [{\citenamefont {Li}\ \emph {et~al.}(2022)\citenamefont {Li},
		\citenamefont {Xue}, \citenamefont {Pan}, \citenamefont {Lin},\ and\
		\citenamefont {Wang}}]{li2022competing}%
	\BibitemOpen
	\bibfield  {author} {\bibinfo {author} {\bibfnamefont {W.}~\bibnamefont
			{Li}}, \bibinfo {author} {\bibfnamefont {X.}~\bibnamefont {Xue}}, \bibinfo
		{author} {\bibfnamefont {L.}~\bibnamefont {Pan}}, \bibinfo {author}
		{\bibfnamefont {T.}~\bibnamefont {Lin}},\ and\ \bibinfo {author}
		{\bibfnamefont {W.}~\bibnamefont {Wang}},\ }\bibfield  {title} {\bibinfo
		{title} {Competing spreading dynamics in simplicial complex},\ }\href
	{https://doi.org/https://doi.org/10.1016/j.amc.2021.126595} {\bibfield
		{journal} {\bibinfo  {journal} {Appl. Math. Comput.}\ }\textbf {\bibinfo
			{volume} {412}},\ \bibinfo {pages} {126595} (\bibinfo {year}
		{2022})}\BibitemShut {NoStop}%
	\bibitem [{\citenamefont {H{\'e}bert-Dufresne}\ \emph
		{et~al.}(2020)\citenamefont {H{\'e}bert-Dufresne}, \citenamefont {Scarpino},\
		and\ \citenamefont {Young}}]{hebert2020macroscopic}%
	\BibitemOpen
	\bibfield  {author} {\bibinfo {author} {\bibfnamefont {L.}~\bibnamefont
			{H{\'e}bert-Dufresne}}, \bibinfo {author} {\bibfnamefont {S.~V.}\
			\bibnamefont {Scarpino}},\ and\ \bibinfo {author} {\bibfnamefont {J.-G.}\
			\bibnamefont {Young}},\ }\bibfield  {title} {\bibinfo {title} {Macroscopic
			patterns of interacting contagions are indistinguishable from social
			reinforcement},\ }\href
	{https://doi.org/https://doi.org/10.1038/s41567-020-0791-2} {\bibfield
		{journal} {\bibinfo  {journal} {Nat. Phys.}\ }\textbf {\bibinfo {volume}
			{16}},\ \bibinfo {pages} {426} (\bibinfo {year} {2020})}\BibitemShut
	{NoStop}%
	\bibitem [{\citenamefont {Centola}\ and\ \citenamefont
		{Macy}(2007)}]{centola2007complex}%
	\BibitemOpen
	\bibfield  {author} {\bibinfo {author} {\bibfnamefont {D.}~\bibnamefont
			{Centola}}\ and\ \bibinfo {author} {\bibfnamefont {M.}~\bibnamefont {Macy}},\
	}\bibfield  {title} {\bibinfo {title} {Complex contagions and the weakness of
			long ties},\ }\href {https://doi.org/https://doi.org/10.1086/521848}
	{\bibfield  {journal} {\bibinfo  {journal} {Am. J. Sociol.}\ }\textbf
		{\bibinfo {volume} {113}},\ \bibinfo {pages} {702} (\bibinfo {year}
		{2007})}\BibitemShut {NoStop}%
	\bibitem [{\citenamefont {Funk}\ \emph {et~al.}(2010)\citenamefont {Funk},
		\citenamefont {Salath\'e},\ and\ \citenamefont {Jansen}}]{funk2010modelling}%
	\BibitemOpen
	\bibfield  {author} {\bibinfo {author} {\bibfnamefont {S.}~\bibnamefont
			{Funk}}, \bibinfo {author} {\bibfnamefont {M.}~\bibnamefont {Salath\'e}},\
		and\ \bibinfo {author} {\bibfnamefont {V.}~\bibnamefont {Jansen}},\
	}\bibfield  {title} {\bibinfo {title} {Modelling the influence of human
			behaviour on the spread of infectious diseases: a review},\ }\href
	{https://doi.org/10.1098/rsif.2010.0142} {\bibfield  {journal} {\bibinfo
			{journal} {J. R. Soc. Interface}\ }\textbf {\bibinfo {volume} {7}},\ \bibinfo
		{pages} {1247} (\bibinfo {year} {2010})}\BibitemShut {NoStop}%
	\bibitem [{\citenamefont {Perra}\ \emph {et~al.}(2011)\citenamefont {Perra},
		\citenamefont {Balcan}, \citenamefont {Gon{\c{c}}alves},\ and\ \citenamefont
		{Vespignani}}]{perra2011towards}%
	\BibitemOpen
	\bibfield  {author} {\bibinfo {author} {\bibfnamefont {N.}~\bibnamefont
			{Perra}}, \bibinfo {author} {\bibfnamefont {D.}~\bibnamefont {Balcan}},
		\bibinfo {author} {\bibfnamefont {B.}~\bibnamefont {Gon{\c{c}}alves}},\ and\
		\bibinfo {author} {\bibfnamefont {A.}~\bibnamefont {Vespignani}},\ }\bibfield
	{title} {\bibinfo {title} {Towards a characterization of behavior-disease
			models},\ }\href
	{https://doi.org/https://doi.org/10.1371/journal.pone.0023084} {\bibfield
		{journal} {\bibinfo  {journal} {PLoS One}\ }\textbf {\bibinfo {volume} {6}},\
		\bibinfo {pages} {e23084} (\bibinfo {year} {2011})}\BibitemShut {NoStop}%
	\bibitem [{\citenamefont {Granell}\ \emph {et~al.}(2013)\citenamefont
		{Granell}, \citenamefont {G\'omez},\ and\ \citenamefont
		{Arenas}}]{granell2013dynamical}%
	\BibitemOpen
	\bibfield  {author} {\bibinfo {author} {\bibfnamefont {C.}~\bibnamefont
			{Granell}}, \bibinfo {author} {\bibfnamefont {S.}~\bibnamefont {G\'omez}},\
		and\ \bibinfo {author} {\bibfnamefont {A.}~\bibnamefont {Arenas}},\
	}\bibfield  {title} {\bibinfo {title} {Dynamical interplay between awareness
			and epidemic spreading in multiplex networks},\ }\href
	{https://doi.org/10.1103/PhysRevLett.111.128701} {\bibfield  {journal}
		{\bibinfo  {journal} {Phys. Rev. Lett.}\ }\textbf {\bibinfo {volume} {111}},\
		\bibinfo {pages} {128701} (\bibinfo {year} {2013})}\BibitemShut {NoStop}%
	\bibitem [{\citenamefont {Scarpino}\ \emph {et~al.}(2016)\citenamefont
		{Scarpino}, \citenamefont {Allard},\ and\ \citenamefont
		{H{\'e}bert-Dufresne}}]{scarpino2016effect}%
	\BibitemOpen
	\bibfield  {author} {\bibinfo {author} {\bibfnamefont {S.~V.}\ \bibnamefont
			{Scarpino}}, \bibinfo {author} {\bibfnamefont {A.}~\bibnamefont {Allard}},\
		and\ \bibinfo {author} {\bibfnamefont {L.}~\bibnamefont
			{H{\'e}bert-Dufresne}},\ }\bibfield  {title} {\bibinfo {title} {The effect of
			a prudent adaptive behaviour on disease transmission},\ }\href
	{https://doi.org/https://doi.org/10.1038/nphys3832} {\bibfield  {journal}
		{\bibinfo  {journal} {Nat. Phys.}\ }\textbf {\bibinfo {volume} {12}},\
		\bibinfo {pages} {1042} (\bibinfo {year} {2016})}\BibitemShut {NoStop}%
	\bibitem [{\citenamefont {Perra}(2021)}]{perra2021non}%
	\BibitemOpen
	\bibfield  {author} {\bibinfo {author} {\bibfnamefont {N.}~\bibnamefont
			{Perra}},\ }\bibfield  {title} {\bibinfo {title} {Non-pharmaceutical
			interventions during the covid-19 pandemic: A review},\ }\href
	{https://doi.org/https://doi.org/10.1016/j.physrep.2021.02.001} {\bibfield
		{journal} {\bibinfo  {journal} {Phys. Rep.}\ }\textbf {\bibinfo {volume}
			{913}},\ \bibinfo {pages} {1} (\bibinfo {year} {2021})}\BibitemShut {NoStop}%
	\bibitem [{\citenamefont {Iacopini}\ \emph {et~al.}(2019)\citenamefont
		{Iacopini}, \citenamefont {Petri}, \citenamefont {Barrat},\ and\
		\citenamefont {Latora}}]{iacopini2019simplicial}%
	\BibitemOpen
	\bibfield  {author} {\bibinfo {author} {\bibfnamefont {I.}~\bibnamefont
			{Iacopini}}, \bibinfo {author} {\bibfnamefont {G.}~\bibnamefont {Petri}},
		\bibinfo {author} {\bibfnamefont {A.}~\bibnamefont {Barrat}},\ and\ \bibinfo
		{author} {\bibfnamefont {V.}~\bibnamefont {Latora}},\ }\bibfield  {title}
	{\bibinfo {title} {Simplicial models of social contagion},\ }\href
	{https://doi.org/https://doi.org/10.1038/s41467-019-10431-6} {\bibfield
		{journal} {\bibinfo  {journal} {Nat. Commun.}\ }\textbf {\bibinfo {volume}
			{10}},\ \bibinfo {pages} {1} (\bibinfo {year} {2019})}\BibitemShut {NoStop}%
	\bibitem [{\citenamefont {Battiston}\ \emph {et~al.}(2020)\citenamefont
		{Battiston}, \citenamefont {Cencetti}, \citenamefont {Iacopini},
		\citenamefont {Latora}, \citenamefont {Lucas}, \citenamefont {Patania},
		\citenamefont {Young},\ and\ \citenamefont {Petri}}]{battiston2020networks}%
	\BibitemOpen
	\bibfield  {author} {\bibinfo {author} {\bibfnamefont {F.}~\bibnamefont
			{Battiston}}, \bibinfo {author} {\bibfnamefont {G.}~\bibnamefont {Cencetti}},
		\bibinfo {author} {\bibfnamefont {I.}~\bibnamefont {Iacopini}}, \bibinfo
		{author} {\bibfnamefont {V.}~\bibnamefont {Latora}}, \bibinfo {author}
		{\bibfnamefont {M.}~\bibnamefont {Lucas}}, \bibinfo {author} {\bibfnamefont
			{A.}~\bibnamefont {Patania}}, \bibinfo {author} {\bibfnamefont {J.-G.}\
			\bibnamefont {Young}},\ and\ \bibinfo {author} {\bibfnamefont
			{G.}~\bibnamefont {Petri}},\ }\bibfield  {title} {\bibinfo {title} {Networks
			beyond pairwise interactions: {{Structure}} and dynamics},\ }\href
	{https://doi.org/10.1016/j.physrep.2020.05.004} {\bibfield  {journal}
		{\bibinfo  {journal} {Phys. Rep.}\ }\textbf {\bibinfo {volume} {874}},\
		\bibinfo {pages} {1} (\bibinfo {year} {2020})}\BibitemShut {NoStop}%
	\bibitem [{\citenamefont {Barrat}\ \emph {et~al.}(2022)\citenamefont {Barrat},
		\citenamefont {Ferraz~de Arruda}, \citenamefont {Iacopini},\ and\
		\citenamefont {Moreno}}]{barrat2022social}%
	\BibitemOpen
	\bibfield  {author} {\bibinfo {author} {\bibfnamefont {A.}~\bibnamefont
			{Barrat}}, \bibinfo {author} {\bibfnamefont {G.}~\bibnamefont {Ferraz~de
				Arruda}}, \bibinfo {author} {\bibfnamefont {I.}~\bibnamefont {Iacopini}},\
		and\ \bibinfo {author} {\bibfnamefont {Y.}~\bibnamefont {Moreno}},\
	}\bibfield  {title} {\bibinfo {title} {Social contagion on higher-order
			structures},\ }in\ \href
	{https://doi.org/https://doi.org/10.1007/978-3-030-91374-8_13} {\emph
		{\bibinfo {booktitle} {Higher-Order Systems}}}\ (\bibinfo  {publisher}
	{Springer},\ \bibinfo {year} {2022})\ pp.\ \bibinfo {pages}
	{329--346}\BibitemShut {NoStop}%
	\bibitem [{\citenamefont {Landry}\ and\ \citenamefont
		{Restrepo}(2020)}]{landry2020effect}%
	\BibitemOpen
	\bibfield  {author} {\bibinfo {author} {\bibfnamefont {N.~W.}\ \bibnamefont
			{Landry}}\ and\ \bibinfo {author} {\bibfnamefont {J.~G.}\ \bibnamefont
			{Restrepo}},\ }\bibfield  {title} {\bibinfo {title} {The effect of
			heterogeneity on hypergraph contagion models},\ }\href
	{https://doi.org/10.1063/5.0020034} {\bibfield  {journal} {\bibinfo
			{journal} {Chaos}\ }\textbf {\bibinfo {volume} {30}},\ \bibinfo {pages}
		{103117} (\bibinfo {year} {2020})}\BibitemShut {NoStop}%
	\bibitem [{\citenamefont {{Ferraz de Arruda}}\ \emph
		{et~al.}(2021)\citenamefont {{Ferraz de Arruda}}, \citenamefont {Tizzani},\
		and\ \citenamefont {Moreno}}]{ferrazdearruda2021phase}%
	\BibitemOpen
	\bibfield  {author} {\bibinfo {author} {\bibfnamefont {G.}~\bibnamefont
			{{Ferraz de Arruda}}}, \bibinfo {author} {\bibfnamefont {M.}~\bibnamefont
			{Tizzani}},\ and\ \bibinfo {author} {\bibfnamefont {Y.}~\bibnamefont
			{Moreno}},\ }\bibfield  {title} {\bibinfo {title} {Phase transitions and
			stability of dynamical processes on hypergraphs},\ }\href
	{https://doi.org/10.1038/s42005-021-00525-3} {\bibfield  {journal} {\bibinfo
			{journal} {Commun. Phys.}\ }\textbf {\bibinfo {volume} {4}},\ \bibinfo
		{pages} {24} (\bibinfo {year} {2021})}\BibitemShut {NoStop}%
	\bibitem [{\citenamefont {Matamalas}\ \emph {et~al.}(2020)\citenamefont
		{Matamalas}, \citenamefont {G{\'o}mez},\ and\ \citenamefont
		{Arenas}}]{matamalas2020abrupt}%
	\BibitemOpen
	\bibfield  {author} {\bibinfo {author} {\bibfnamefont {J.~T.}\ \bibnamefont
			{Matamalas}}, \bibinfo {author} {\bibfnamefont {S.}~\bibnamefont
			{G{\'o}mez}},\ and\ \bibinfo {author} {\bibfnamefont {A.}~\bibnamefont
			{Arenas}},\ }\bibfield  {title} {\bibinfo {title} {Abrupt phase transition of
			epidemic spreading in simplicial complexes},\ }\href
	{https://doi.org/10.1103/PhysRevResearch.2.012049} {\bibfield  {journal}
		{\bibinfo  {journal} {Phys. Rev. Research}\ }\textbf {\bibinfo {volume}
			{2}},\ \bibinfo {pages} {012049} (\bibinfo {year} {2020})}\BibitemShut
	{NoStop}%
	\bibitem [{\citenamefont {St-Onge}\ \emph {et~al.}(2022)\citenamefont
		{St-Onge}, \citenamefont {Iacopini}, \citenamefont {Latora}, \citenamefont
		{Barrat}, \citenamefont {Petri}, \citenamefont {Allard},\ and\ \citenamefont
		{H{\'e}bert-Dufresne}}]{st2022influential}%
	\BibitemOpen
	\bibfield  {author} {\bibinfo {author} {\bibfnamefont {G.}~\bibnamefont
			{St-Onge}}, \bibinfo {author} {\bibfnamefont {I.}~\bibnamefont {Iacopini}},
		\bibinfo {author} {\bibfnamefont {V.}~\bibnamefont {Latora}}, \bibinfo
		{author} {\bibfnamefont {A.}~\bibnamefont {Barrat}}, \bibinfo {author}
		{\bibfnamefont {G.}~\bibnamefont {Petri}}, \bibinfo {author} {\bibfnamefont
			{A.}~\bibnamefont {Allard}},\ and\ \bibinfo {author} {\bibfnamefont
			{L.}~\bibnamefont {H{\'e}bert-Dufresne}},\ }\bibfield  {title} {\bibinfo
		{title} {Influential groups for seeding and sustaining nonlinear contagion in
			heterogeneous hypergraphs},\ }\href
	{https://doi.org/https://doi.org/10.1038/s42005-021-00788-w} {\bibfield
		{journal} {\bibinfo  {journal} {Commun. Phys.}\ }\textbf {\bibinfo {volume}
			{5}},\ \bibinfo {pages} {1} (\bibinfo {year} {2022})}\BibitemShut {NoStop}%
	\bibitem [{\citenamefont {Ventura}\ \emph {et~al.}(2021)\citenamefont
		{Ventura}, \citenamefont {Moreno},\ and\ \citenamefont
		{Rodrigues}}]{ventura2021role}%
	\BibitemOpen
	\bibfield  {author} {\bibinfo {author} {\bibfnamefont {P.~C.}\ \bibnamefont
			{Ventura}}, \bibinfo {author} {\bibfnamefont {Y.}~\bibnamefont {Moreno}},\
		and\ \bibinfo {author} {\bibfnamefont {F.~A.}\ \bibnamefont {Rodrigues}},\
	}\bibfield  {title} {\bibinfo {title} {Role of time scale in the spreading of
			asymmetrically interacting diseases},\ }\href
	{https://doi.org/10.1103/PhysRevResearch.3.013146} {\bibfield  {journal}
		{\bibinfo  {journal} {Phys. Rev. Research}\ }\textbf {\bibinfo {volume}
			{3}},\ \bibinfo {pages} {013146} (\bibinfo {year} {2021})}\BibitemShut
	{NoStop}%
	\bibitem [{\citenamefont {Lambiotte}\ \emph {et~al.}(2019)\citenamefont
		{Lambiotte}, \citenamefont {Rosvall},\ and\ \citenamefont
		{Scholtes}}]{lambiotte2019networks}%
	\BibitemOpen
	\bibfield  {author} {\bibinfo {author} {\bibfnamefont {R.}~\bibnamefont
			{Lambiotte}}, \bibinfo {author} {\bibfnamefont {M.}~\bibnamefont {Rosvall}},\
		and\ \bibinfo {author} {\bibfnamefont {I.}~\bibnamefont {Scholtes}},\
	}\bibfield  {title} {\bibinfo {title} {From networks to optimal higher-order
			models of complex systems},\ }\bibfield  {journal} {\bibinfo  {journal} {Nat.
			Phys.}\ }\href {https://doi.org/10.1038/s41567-019-0459-y}
	{10.1038/s41567-019-0459-y} (\bibinfo {year} {2019})\BibitemShut {NoStop}%
	\bibitem [{\citenamefont {Bick}\ \emph {et~al.}(2021)\citenamefont {Bick},
		\citenamefont {Gross}, \citenamefont {Harrington},\ and\ \citenamefont
		{Schaub}}]{bick2021higher}%
	\BibitemOpen
	\bibfield  {author} {\bibinfo {author} {\bibfnamefont {C.}~\bibnamefont
			{Bick}}, \bibinfo {author} {\bibfnamefont {E.}~\bibnamefont {Gross}},
		\bibinfo {author} {\bibfnamefont {H.~A.}\ \bibnamefont {Harrington}},\ and\
		\bibinfo {author} {\bibfnamefont {M.~T.}\ \bibnamefont {Schaub}},\ }\bibfield
	{title} {\bibinfo {title} {What are higher-order networks?},\ }\bibfield
	{journal} {\bibinfo  {journal} {arXiv preprint arXiv:2104.11329}\ }\href
	{https://doi.org/https://doi.org/10.48550/arXiv.2104.11329}
	{https://doi.org/10.48550/arXiv.2104.11329} (\bibinfo {year}
	{2021})\BibitemShut {NoStop}%
	\bibitem [{\citenamefont {Battiston}\ \emph {et~al.}(2021)\citenamefont
		{Battiston}, \citenamefont {Amico}, \citenamefont {Barrat}, \citenamefont
		{Bianconi}, \citenamefont {Ferraz~de Arruda}, \citenamefont {Franceschiello},
		\citenamefont {Iacopini}, \citenamefont {K{\'e}fi}, \citenamefont {Latora},
		\citenamefont {Moreno}, \citenamefont {Murray}, \citenamefont {Peixoto},
		\citenamefont {Vaccarino},\ and\ \citenamefont
		{Petri}}]{battiston2021physics}%
	\BibitemOpen
	\bibfield  {author} {\bibinfo {author} {\bibfnamefont {F.}~\bibnamefont
			{Battiston}}, \bibinfo {author} {\bibfnamefont {E.}~\bibnamefont {Amico}},
		\bibinfo {author} {\bibfnamefont {A.}~\bibnamefont {Barrat}}, \bibinfo
		{author} {\bibfnamefont {G.}~\bibnamefont {Bianconi}}, \bibinfo {author}
		{\bibfnamefont {G.}~\bibnamefont {Ferraz~de Arruda}}, \bibinfo {author}
		{\bibfnamefont {B.}~\bibnamefont {Franceschiello}}, \bibinfo {author}
		{\bibfnamefont {I.}~\bibnamefont {Iacopini}}, \bibinfo {author}
		{\bibfnamefont {S.}~\bibnamefont {K{\'e}fi}}, \bibinfo {author}
		{\bibfnamefont {V.}~\bibnamefont {Latora}}, \bibinfo {author} {\bibfnamefont
			{Y.}~\bibnamefont {Moreno}}, \bibinfo {author} {\bibfnamefont
			{M.}~\bibnamefont {Murray}}, \bibinfo {author} {\bibfnamefont
			{T.}~\bibnamefont {Peixoto}}, \bibinfo {author} {\bibfnamefont
			{F.}~\bibnamefont {Vaccarino}},\ and\ \bibinfo {author} {\bibfnamefont
			{G.}~\bibnamefont {Petri}},\ }\bibfield  {title} {\bibinfo {title} {The
			physics of higher-order interactions in complex systems},\ }\href
	{https://doi.org/10.1038/s41567-021-01371-4} {\bibfield  {journal} {\bibinfo
			{journal} {Nat. Phys.}\ }\textbf {\bibinfo {volume} {17}},\ \bibinfo {pages}
		{1093} (\bibinfo {year} {2021})}\BibitemShut {NoStop}%
	\bibitem [{\citenamefont {Torres}\ \emph {et~al.}(2021)\citenamefont {Torres},
		\citenamefont {Blevins}, \citenamefont {Bassett},\ and\ \citenamefont
		{Eliassi-Rad}}]{torres2021and}%
	\BibitemOpen
	\bibfield  {author} {\bibinfo {author} {\bibfnamefont {L.}~\bibnamefont
			{Torres}}, \bibinfo {author} {\bibfnamefont {A.~S.}\ \bibnamefont {Blevins}},
		\bibinfo {author} {\bibfnamefont {D.}~\bibnamefont {Bassett}},\ and\ \bibinfo
		{author} {\bibfnamefont {T.}~\bibnamefont {Eliassi-Rad}},\ }\bibfield
	{title} {\bibinfo {title} {The why, how, and when of representations for
			complex systems},\ }\href
	{https://doi.org/https://doi.org/10.1137/20M1355896} {\bibfield  {journal}
		{\bibinfo  {journal} {SIAM Rev.}\ }\textbf {\bibinfo {volume} {63}},\
		\bibinfo {pages} {435} (\bibinfo {year} {2021})}\BibitemShut {NoStop}%
	\bibitem [{\citenamefont {Kiss}\ \emph {et~al.}(2017)\citenamefont {Kiss},
		\citenamefont {Miller},\ and\ \citenamefont {Simon}}]{kiss2017mathematics}%
	\BibitemOpen
	\bibfield  {author} {\bibinfo {author} {\bibfnamefont {I.}~\bibnamefont
			{Kiss}}, \bibinfo {author} {\bibfnamefont {J.}~\bibnamefont {Miller}},\ and\
		\bibinfo {author} {\bibfnamefont {P.}~\bibnamefont {Simon}},\ }\href
	{https://books.google.at/books?id=DlEnDwAAQBAJ} {\emph {\bibinfo {title}
			{Mathematics of Epidemics on Networks: From Exact to Approximate Models}}},\
	Interdisciplinary Applied Mathematics\ (\bibinfo  {publisher} {Springer
		International Publishing},\ \bibinfo {year} {2017})\BibitemShut {NoStop}%
	\bibitem [{\citenamefont {G{\'o}mez}\ \emph {et~al.}(2010)\citenamefont
		{G{\'o}mez}, \citenamefont {Arenas}, \citenamefont {Borge-Holthoefer},
		\citenamefont {Meloni},\ and\ \citenamefont {Moreno}}]{gomez2010discrete}%
	\BibitemOpen
	\bibfield  {author} {\bibinfo {author} {\bibfnamefont {S.}~\bibnamefont
			{G{\'o}mez}}, \bibinfo {author} {\bibfnamefont {A.}~\bibnamefont {Arenas}},
		\bibinfo {author} {\bibfnamefont {J.}~\bibnamefont {Borge-Holthoefer}},
		\bibinfo {author} {\bibfnamefont {S.}~\bibnamefont {Meloni}},\ and\ \bibinfo
		{author} {\bibfnamefont {Y.}~\bibnamefont {Moreno}},\ }\bibfield  {title}
	{\bibinfo {title} {Discrete-time markov chain approach to contact-based
			disease spreading in complex networks},\ }\href
	{https://doi.org/https://doi.org/10.1209/0295-5075/89/38009} {\bibfield
		{journal} {\bibinfo  {journal} {Europhys. Lett.}\ }\textbf {\bibinfo {volume}
			{89}},\ \bibinfo {pages} {38009} (\bibinfo {year} {2010})}\BibitemShut
	{NoStop}%
	\bibitem [{\citenamefont {Soriano-Pa{\~n}os}\ \emph {et~al.}(2019)\citenamefont
		{Soriano-Pa{\~n}os}, \citenamefont {Ghanbarnejad}, \citenamefont {Meloni},\
		and\ \citenamefont {G{\'o}mez-Garde{\~n}es}}]{soriano2019markovian}%
	\BibitemOpen
	\bibfield  {author} {\bibinfo {author} {\bibfnamefont {D.}~\bibnamefont
			{Soriano-Pa{\~n}os}}, \bibinfo {author} {\bibfnamefont {F.}~\bibnamefont
			{Ghanbarnejad}}, \bibinfo {author} {\bibfnamefont {S.}~\bibnamefont
			{Meloni}},\ and\ \bibinfo {author} {\bibfnamefont {J.}~\bibnamefont
			{G{\'o}mez-Garde{\~n}es}},\ }\bibfield  {title} {\bibinfo {title} {Markovian
			approach to tackle the interaction of simultaneous diseases},\ }\href
	{https://doi.org/https://doi.org/10.1103/PhysRevE.100.062308} {\bibfield
		{journal} {\bibinfo  {journal} {Phys. Rev. E}\ }\textbf {\bibinfo {volume}
			{100}},\ \bibinfo {pages} {062308} (\bibinfo {year} {2019})}\BibitemShut
	{NoStop}%
	\bibitem [{\citenamefont {G{\'o}mez-Gardenes}\ \emph
		{et~al.}(2016)\citenamefont {G{\'o}mez-Gardenes}, \citenamefont {Lotero},
		\citenamefont {Taraskin},\ and\ \citenamefont
		{P{\'e}rez-Reche}}]{gomez2016explosive}%
	\BibitemOpen
	\bibfield  {author} {\bibinfo {author} {\bibfnamefont {J.}~\bibnamefont
			{G{\'o}mez-Gardenes}}, \bibinfo {author} {\bibfnamefont {L.}~\bibnamefont
			{Lotero}}, \bibinfo {author} {\bibfnamefont {S.}~\bibnamefont {Taraskin}},\
		and\ \bibinfo {author} {\bibfnamefont {F.}~\bibnamefont {P{\'e}rez-Reche}},\
	}\bibfield  {title} {\bibinfo {title} {Explosive contagion in networks},\
	}\href {https://doi.org/https://doi.org/10.1038/srep19767} {\bibfield
		{journal} {\bibinfo  {journal} {Sci. Rep.}\ }\textbf {\bibinfo {volume}
			{6}},\ \bibinfo {pages} {1} (\bibinfo {year} {2016})}\BibitemShut {NoStop}%
	\bibitem [{\citenamefont {Kuehn}\ and\ \citenamefont
		{Bick}(2021)}]{kuehn2021universal}%
	\BibitemOpen
	\bibfield  {author} {\bibinfo {author} {\bibfnamefont {C.}~\bibnamefont
			{Kuehn}}\ and\ \bibinfo {author} {\bibfnamefont {C.}~\bibnamefont {Bick}},\
	}\bibfield  {title} {\bibinfo {title} {A universal route to explosive
			phenomena},\ }\href {https://doi.org/10.1126/sciadv.abe3824} {\bibfield
		{journal} {\bibinfo  {journal} {Sci. Adv.}\ }\textbf {\bibinfo {volume}
			{7}},\ \bibinfo {pages} {eabe3824} (\bibinfo {year} {2021})}\BibitemShut
	{NoStop}%
	\bibitem [{\citenamefont {Khanra}\ and\ \citenamefont
		{Pal}(2021)}]{khanra2021explosive}%
	\BibitemOpen
	\bibfield  {author} {\bibinfo {author} {\bibfnamefont {P.}~\bibnamefont
			{Khanra}}\ and\ \bibinfo {author} {\bibfnamefont {P.}~\bibnamefont {Pal}},\
	}\bibfield  {title} {\bibinfo {title} {Explosive synchronization in
			multilayer networks through partial adaptation},\ }\href
	{https://doi.org/10.1016/j.chaos.2020.110621} {\bibfield  {journal} {\bibinfo
			{journal} {Chaos Solit. Fractals}\ }\textbf {\bibinfo {volume} {143}},\
		\bibinfo {pages} {110621} (\bibinfo {year} {2021})}\BibitemShut {NoStop}%
	\bibitem [{\citenamefont {Brett}\ \emph {et~al.}(2020)\citenamefont {Brett},
		\citenamefont {Ajelli}, \citenamefont {Liu}, \citenamefont {Krauland},
		\citenamefont {Grefenstette}, \citenamefont {van Panhuis}, \citenamefont
		{Vespignani}, \citenamefont {Drake},\ and\ \citenamefont
		{Rohani}}]{brett2020detecting}%
	\BibitemOpen
	\bibfield  {author} {\bibinfo {author} {\bibfnamefont {T.}~\bibnamefont
			{Brett}}, \bibinfo {author} {\bibfnamefont {M.}~\bibnamefont {Ajelli}},
		\bibinfo {author} {\bibfnamefont {Q.-H.}\ \bibnamefont {Liu}}, \bibinfo
		{author} {\bibfnamefont {M.~G.}\ \bibnamefont {Krauland}}, \bibinfo {author}
		{\bibfnamefont {J.~J.}\ \bibnamefont {Grefenstette}}, \bibinfo {author}
		{\bibfnamefont {W.~G.}\ \bibnamefont {van Panhuis}}, \bibinfo {author}
		{\bibfnamefont {A.}~\bibnamefont {Vespignani}}, \bibinfo {author}
		{\bibfnamefont {J.~M.}\ \bibnamefont {Drake}},\ and\ \bibinfo {author}
		{\bibfnamefont {P.}~\bibnamefont {Rohani}},\ }\bibfield  {title} {\bibinfo
		{title} {Detecting critical slowing down in high-dimensional epidemiological
			systems},\ }\href {https://doi.org/10.1371/journal.pcbi.1007679} {\bibfield
		{journal} {\bibinfo  {journal} {PLoS Comput. Biol.}\ }\textbf {\bibinfo
			{volume} {16}},\ \bibinfo {pages} {e1007679} (\bibinfo {year}
		{2020})}\BibitemShut {NoStop}%
	\bibitem [{\citenamefont {Peixoto}(2019)}]{peixoto2019network}%
	\BibitemOpen
	\bibfield  {author} {\bibinfo {author} {\bibfnamefont {T.~P.}\ \bibnamefont
			{Peixoto}},\ }\bibfield  {title} {\bibinfo {title} {Network reconstruction
			and community detection from dynamics},\ }\href
	{https://doi.org/10.1103/PhysRevLett.123.128301} {\bibfield  {journal}
		{\bibinfo  {journal} {Phys. Rev. Lett.}\ }\textbf {\bibinfo {volume} {123}},\
		\bibinfo {pages} {128301} (\bibinfo {year} {2019})}\BibitemShut {NoStop}%
	\bibitem [{\citenamefont {Inc.}()}]{Mathematica}%
	\BibitemOpen
	\bibfield  {author} {\bibinfo {author} {\bibfnamefont {W.~R.}\ \bibnamefont
			{Inc.}},\ }\href {https://www.wolfram.com/mathematica} {\bibinfo {title}
		{Mathematica, {V}ersion 13.3.0.0}},\ \bibinfo {note} {champaign, IL,
		2022}\BibitemShut {NoStop}%
\end{thebibliography}

%apsrev4-2.bst 2019-01-14 (MD) hand-edited version of apsrev4-1.bst
%Control: key (0)
%Control: author (8) initials jnrlst
%Control: editor formatted (1) identically to author
%Control: production of article title (0) allowed
%Control: page (0) single
%Control: year (1) truncated
%Control: production of eprint (0) enabled
%

%%%%%%%%%%%%%%%%%%%%%%%%%%%%%%%%%%%%%%%%%%%%%%%%%%%%%%%%%%%%%%%%
% SUPPLEMENTARY MATERIAL
%%%%%%%%%%%%%%%%%%%%%%%%%%%%%%%%%%%%%%%%%%%%%%%%%%%%%%%%%%%%%%%%
\clearpage
\newpage

%%%%%%%%%% Prefix a "S" to all equations, figures, equations, tables and reset the counter %%%%%%%%%%
\setcounter{figure}{0}
\setcounter{table}{0}
\setcounter{equation}{0}
\setcounter{page}{1}
\setcounter{section}{0}

\makeatletter
\renewcommand{\thefigure}{S\arabic{figure}}
\renewcommand{\theequation}{S\arabic{equation}}
\renewcommand{\thetable}{S\arabic{table}}
%%%%%%%%%% Prefix a "S" to all equations, figures, equations, tables and reset the counter %%%%%%%%%%

\setcounter{secnumdepth}{2}

\widetext
\begin{center}
	\textbf{\large Supplementary Material: Simplicially driven simple contagion}
\end{center}

\section{Case of different recovery rates: $\mu_A \neq \mu_B$}
\label{SI:sec:mua_mub}
In the main text we assumed identical recovery rates. Here, we remove this constraint and allow them to be potentially different, so that $\mu_A \neq \mu_B$.
By rescaling all equations by $\mu_A$ (instead of $\mu$), we have the following---instead of Eqs. \eqref{SI:eq:MF_driven}:
\begin{subequations}
	\begin{align}
\dot \rho_A =& - 1 \rho_A + \lambda_A \rho_S (\rho_A + \rho_{AB})
+ \lambda_A^{\triangle} \rho_S (\rho_A + \rho_{AB})^2 %\nonumber \\
+ \frac{\mu_B }{\mu_A} \rho_{AB} - \epsAB \lambda_B \frac{\mu_B }{\mu_A} \rho_A (\rho_B + \rho_{AB})  \\
\dot \rho_B =& - \frac{\mu_B }{\mu_A} \rho_B + \lambda_B \frac{\mu_B }{\mu_A} \rho_S (\rho_B + \rho_{AB}) %\nonumber \\
+ 1 \rho_{AB} - \lambda_A \rho_B (\rho_A + \rho_{AB}) %\nonumber \\
- \lambda_A^{\triangle} \rho_B (\rho_A + \rho_{AB})^2  \\
\dot \rho_{AB} =& - (1 + \frac{\mu_B }{\mu_A}) \rho_{AB}+ \epsAB \lambda_B \frac{\mu_B }{\mu_A} \rho_A (\rho_B + \rho_{AB}) %\nonumber \\
+ \lambda_A \rho_B (\rho_A + \rho_{AB}) %\nonumber\\ 
+ \lambda_A^{\triangle} \rho_B (\rho_A + \rho_{AB})^2      
\end{align}
\end{subequations}
which, introducing the total densities, becomes
\begin{subequations}
\begin{align}
\dot \rho_{A_{\text{tot}}} ={}&  (- \rho_A - \rho_{AB}) + \lambda_A \rho_{A_{\text{tot}}} [1 - \rho_{B_{\text{tot}}} - \rho_A + \rho_B]  + \lambda_A^{\triangle} \rho_{A_{\text{tot}}}^2  [1 - \rho_{B_{\text{tot}}} - \rho_A + \rho_B]  \\
\dot  \rho_{B_{\text{tot}}} =&  (- \rho_B - \rho_{AB}) \frac{\mu_B }{\mu_A}  + \lambda_B \frac{\mu_B }{\mu_A} \rho_{B_{\text{tot}}} [1 - \rho_{A_{\text{tot}}} - \rho_B + \epsAB \rho_A]  \\		
\dot \rho_{AB} ={}& - (1 + \frac{\mu_B }{\mu_A}) \rho_{AB}+ \epsAB \lambda_B \frac{\mu_B }{\mu_A} \rho_A \rho_{B_{\text{tot}}} +  \lambda_A \rho_B \rho_{A_{\text{tot}}} +  \lambda_A^{\triangle} \rho_B \rho_{A_{\text{tot}}}^2 .
\end{align}
\end{subequations}
and then
\begin{subequations}
\begin{align}
\dot \rho_{A_{\text{tot}}} ={}& - \rho_{A_{\text{tot}}} 1  + \lambda_A \rho_{A_{\text{tot}}} [1 - \rho_{A_{\text{tot}}}]  + \lambda_A^{\triangle} \rho_{A_{\text{tot}}}^2  [1 - \rho_{A_{\text{tot}}}]  \\
\dot  \rho_{B_{\text{tot}}} =&  - \rho_{B_{\text{tot}}} \frac{\mu_B }{\mu_A}  + \lambda_B \frac{\mu_B }{\mu_A} \rho_{B_{\text{tot}}} [1 - \rho_{A_{\text{tot}}} - \rho_{B_{\text{tot}}} +  \rho_{AB} + \epsAB (\rho_{A_{\text{tot}}} - \rho_{AB})]  \\		
\dot \rho_{AB} ={}& - (1 + \frac{\mu_B }{\mu_A}) \rho_{AB}+ \epsAB \lambda_B \frac{\mu_B }{\mu_A} (\rho_{A_{\text{tot}}} - \rho_{AB}) \rho_{B_{\text{tot}}} +  \lambda_A (\rho_{B_{\text{tot}}} - \rho_{AB}) \rho_{A_{\text{tot}}} +  \lambda_A^{\triangle} (\rho_{B_{\text{tot}}} - \rho_{AB}) \rho_{A_{\text{tot}}}^2
\end{align}
\end{subequations}
which, compared to the case of identical recovery rates, contain the additional $\frac{\mu_B }{\mu_A}$ factors.
We denote that dimensionless ratio $\delta = \frac{\mu_B }{\mu_A}$ 
and the equations become, after refactoring:
\begin{subequations} 
\begin{align}
\dot \rho_{A_{\text{tot}}} &= \rho_{A_{\text{tot}}} \left[ -1 + \lambda_A (1 - \rho_{A_{\text{tot}}}) \right. 
 + \lambda_A^{\triangle} \rho_{A_{\text{tot}}}  (1 - \rho_{A_{\text{tot}}}) ]  , \\
\dot \rho_{B_{\text{tot}}} &= \rho_{B_{\text{tot}}} \delta  \left[ - 1 + \lambda_B (1 - \rho_{B_{\text{tot}}}) \right.  
  \left. + \lambda_B (\epsAB - 1 ) (\rho_{A_{\text{tot}}} -  \rho_{AB})  \right] ,  \\
\dot \rho_{AB} &= - (1 + \delta) \rho_{AB}+ \epsAB \lambda_B \delta  (\rho_{A_{\text{tot}}} - \rho_{AB}) \rho_{B_{\text{tot}}} 
 +  \lambda_A (\rho_{B_{\text{tot}}} - \rho_{AB}) \rho_{A_{\text{tot}}} +  \lambda_A^{\triangle} (\rho_{B_{\text{tot}}} - \rho_{AB}) \rho_{A_{\text{tot}}}^2 .
\end{align} 
\end{subequations}
So, the equation for $\rho_{A_{\text{tot}}}$ (simplagion) is unchanged, as expected. For $\rho_{B_{\text{tot}}}$, we notice a temporal rescaling by a factor $\delta$, but the implicit solution is unchanged,
\begin{equation}
\rho_{B_{\text{tot}}}^{*, \pm} = 1 - \frac{1 }{\lambda_B} + (\rho_{A_{\text{tot}}}^{*, \pm} - \rho_{AB}^{*, \pm})  (\epsAB - 1 ).
\end{equation}

We can consider two limits where the timescales for $A$ and $B$ are of different orders. First, in the limit $\delta \ll 1$, which means that $B$ heals much slower than $A$, $\dot \rho_{B_{\text{tot}}} \approx 0$, that is process $B$ is quasi-static compared to the timescale of process $A$. Thus, $\rho_{A_{\text{tot}}}$ converges fast to its NESS and $\rho_{B_{\text{tot}}}$ is driven by that NESS. Second, in the limit $\delta \gg 1$, $B$ heals much faster than $A$, it is the opposite. It is possible then to rescale time by $\delta$ to see that process $A$ now appears quasi-static compared to the timescale of $B$. So, $\rho_{B_{\text{tot}}}$ converges fast to its NESS which is in fact adiabatically moving towards its asymptotic NESS, driven by $\rho_{A_{\text{tot}}}$ that slowly converges to its own NESS.

\clearpage

\begin{figure}[b]
    \centering
\includegraphics[width=\linewidth]{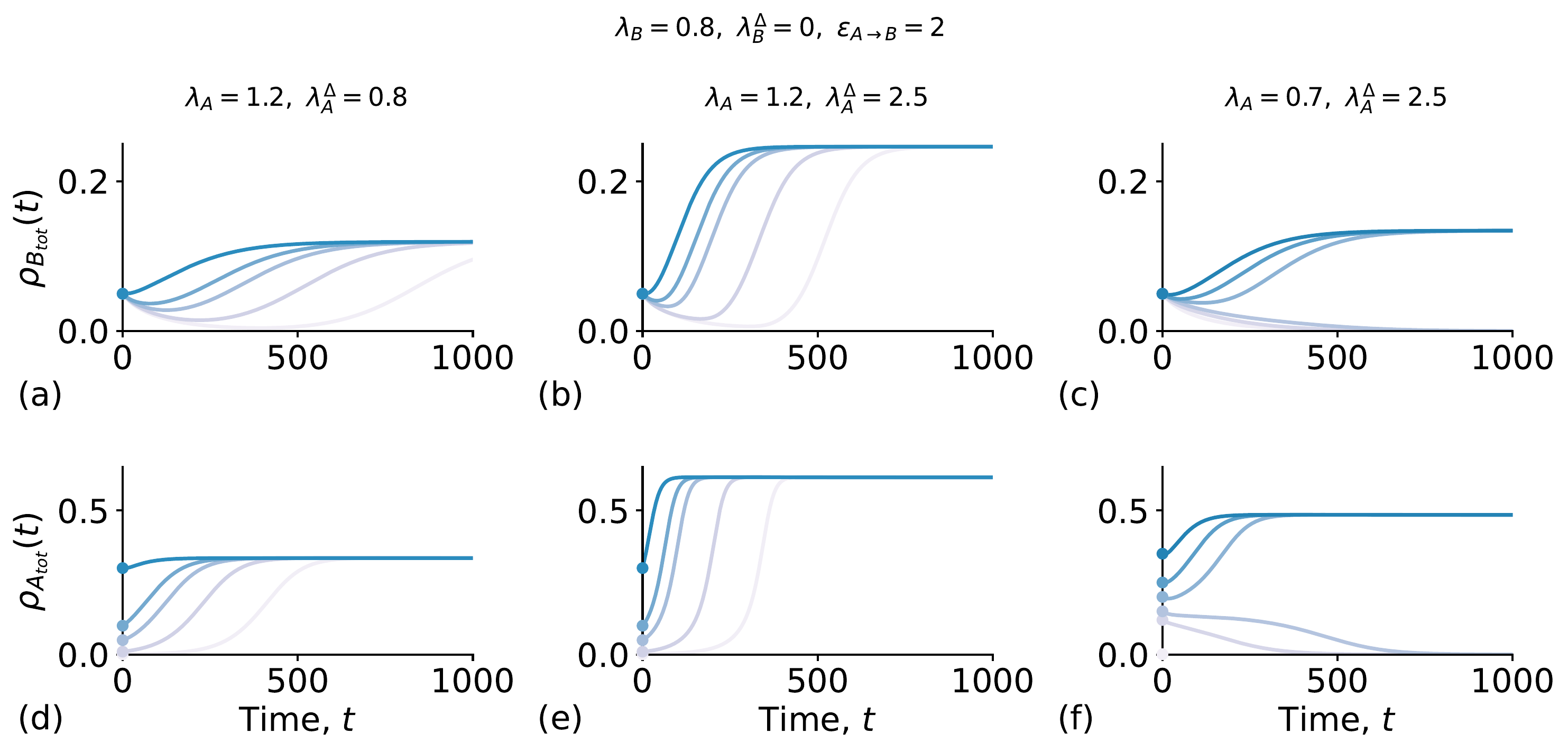}
        \caption{The temporal evolution of the simple contagion $B$ is affected by the initial conditions of the hidden driver process $A$. As for Fig.~5 of the main text, we show $\rho_{B_{\text{tot}}}$ over time (a-c), but together with the temporal dynamics of the driver process, as given by $\rho_{A_{\text{tot}}}$ (d-f). In (a,d) a simple driver process is used ($\lambda_A^{\triangle}=0.8$), while in (b,e) and (c,f) the driver process $A$ is truly simplicial ($\lambda_A^{\triangle}=2.5$). The process $A$ is placed either in the endemic region, $\lambda_A=1.2$ [(a,d) and (b,e)] or in the bi-stable region ($\lambda_A=0.7$). Different curves correspond to different initial conditions of the driver process, $\rho_A(0)$. The other parameters are set to $\lambda_B=0.8$, $\lambda_B^{\triangle}=0$, and $\epsAB=2$.}
    \label{SI:fig:temporal_dynamics}
\end{figure}

\begin{figure}[b]
    \centering
\includegraphics[width=0.4\linewidth]{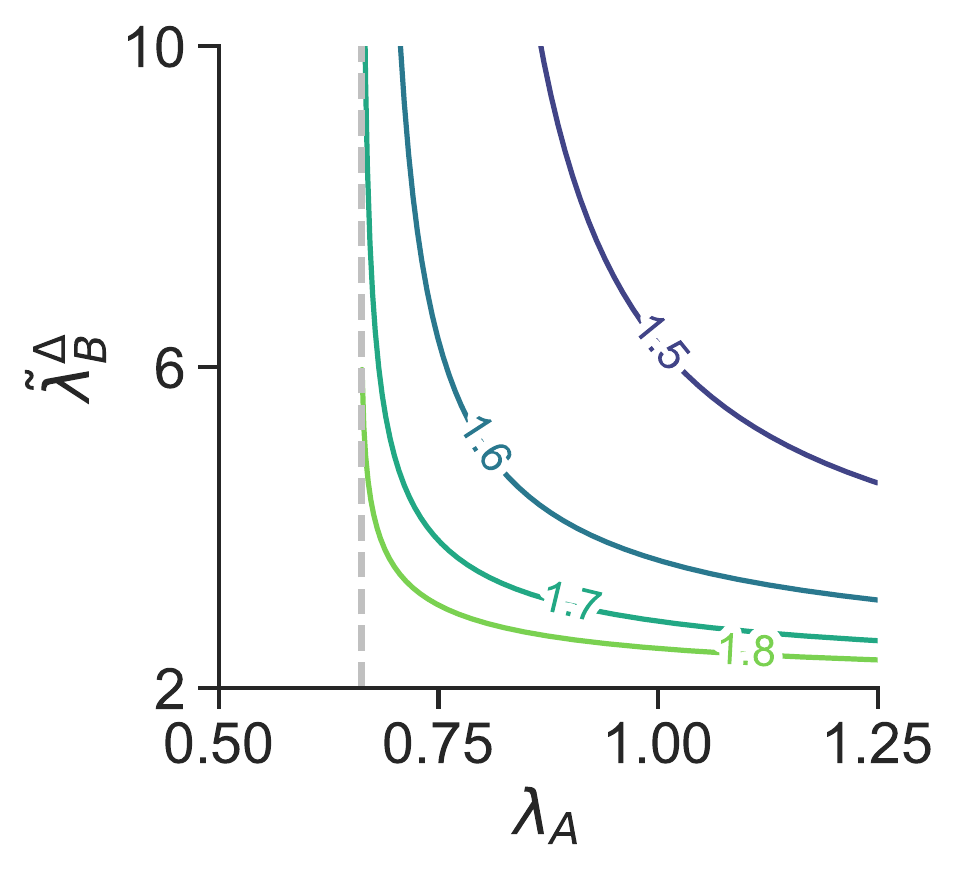}
        \caption{Effective triangle infectivity $\tilde \lambda_B^{\triangle}$ of simple contagion $B$ as a function of $\lambda_A$, for several values of the interaction $\epsAB$ (indicated on the curves). The dashed grey curve indicates the value $\lambda_A^c$, where  diverge $\tilde \lambda_B^{\triangle}$ diverges.}
    \label{fig:effective_triangle}
\end{figure}

\end{document}